\documentclass[aps,prb,twocolumn,preprintnumber]{revtex4-1}
\usepackage{graphicx}
\usepackage{amsmath}
\usepackage{bm}
\usepackage{color}
\usepackage{epstopdf}

\begin{document}

\title{Site dependency of the magnetism for Mn adsorption on MgO/Ag(001): a combined DFT$+U$ and STM investigation}
\author{Shruba Gangopadhyay$^{1,2}$, Thaneshwor P. Kaloni$^3$, Udo Schwingenschl\"ogl$^3$, Barbara A. Jones$^1$}

\affiliation{$^1$IBM Research - Almaden, 650 Harry Road, San Jose, CA 95120}
\affiliation{$^2$ Department of Physics, University of California, Davis CA 95616, USA}
\affiliation{$^3$PSE Division, KAUST, Thuwal 23955-6900, Kingdom of Saudi Arabia}

\begin{abstract}
Theoretical and experimental investigation of the electronic and magnetic structure of transition metal atoms on an insulating interface with a metallic substrate at low temperatures is quite challenging. In this paper, we show a density functional theory plus Hubbard $U$ based protocol to study an Mn adatom on three symmetrically allowed absorption sites, namely on O, hollow (between two Mg and two O) and on Mg on MgO(001). We added a thick enough(bulk like) metallic Ag slab beneath MgO for faithful replication of scanning tunneling microscopy (STM) experiments. Our study reveals that to determine the stable most binding site, we need to obtain a Hubbard $U$ value from the density functional theory(DFT) calculations, and our results show agreement with STM experiments. Our calculated Hubbard $U$ values for the three adatom sites are different. When Mn sits on O, it retains 2.4 $\mu_B$ spin moment, close to its atomic spin moment. However, when Mn sits on other adatom sites, the spin moment decreases.  Using the atom projected density of states, we find Mn on O atom shows very narrow crystal field splitting among \textit{d} orbitals, whereas on other adatom sites Mn \textit{d} orbitals show significant splitting. We calculate charge and spin densities and show vertical and horizontal propagation of charge and spin density varies widely between sites. Mn on Mg top shows an unusual feature, that Mn pushes Mg below, to make a coplanar of Mn geometry with the four oxygen atoms. In addition we explained the reason of spin leaking down to the Ag layers.  Mn on MgO/Ag does not show any spin-flip behavior in the STM, an unusual phenomenon, and we used our first principles-based approach to explain this unique observation.
\end{abstract}
\date{\today}

\maketitle
\noindent \textbf{Introduction}\\
Miniaturization of electronic device interfaces between metals and oxides is interesting for tuning materials properties. For example, oxides are used as substrates for metallic nanoparticles.\cite{Freund} Thin oxide films on metallic surfaces form hybrid systems with various applications such as in electronic devices \cite{Robertson} and chemical sensors.\cite{Netzer} In general, the interaction between the oxide and metal can strongly affect the structural and electronic properties of oxide films, and this determines the device- performance. The effect of the work function of the metallic substrate has been studied widely by both experiment\cite{Greiner,Jaouen} and theory.\cite{Giordano, Prada} In microelectronic devices, for instance, MgO is often used as an insulating barrier.\cite{Brongersma} 

Considerable efforts have been devoted to understanding the properties of metallic clusters on bulk oxide surfaces\cite{Chen,Glaspell} as well as on thin oxide films.\cite{Luo,Harding,Lin} For thin-film MgO supported by Ag(001), the effect of structural relaxation \cite{Valeri,Ferrari,Bieletzki,Noguera}, the interface properties,\cite{Prada} the electronic density of states (DOS) \cite{Schintke}, and various defects \cite{Sterrer} have been studied. The most common characterization techniques are STM, low-energy electron diffraction, atomic emission spectroscopy, and ultraviolet photoelectron spectroscopy to obtain insight into such systems.\cite{Cabailh,Ouvrard} Some recent experimental studies on a bare MgO/Ag(001) surface examined by STM \cite{acsnano}, and related density functional theory (DFT) study were also reported.\cite{beigi2014} But this DFT analysis underplayed the effect of the Ag layers and used very few (insufficient in our view) metallic layers that are required for this type of analysis, fewer metal substrate layers yield significant dipole correction error which leads to incorrect description of electronic structure.

Experimentally and theoretically, the structural and electronic properties as well as charge redistribution have been studied for Cr doped MgO and Mo doped CaO.\cite{Stavale} Though the surfaces of MgO and CaO are isostructural and isoelectronic and the dopants Cr and Mo exhibit a similar electronic behavior (including the ionization energies), Cr doped MgO and Mo doped CaO behave very differently. This confirms that the structural, electronic, and magnetic properties of doped MgO are not well understood. Taking advantage of the versatility of first-principles calculations, we therefore study in the present work Mn-adatoms on MgO(001)/Ag(001). We will argue that depending on the binding site the magnetic properties are very different and that therefore the system can be flexibly tuned, which is desirable for magnetic memory devices, for example.

\begin{figure}[ht]
\includegraphics[width=0.5\textwidth]{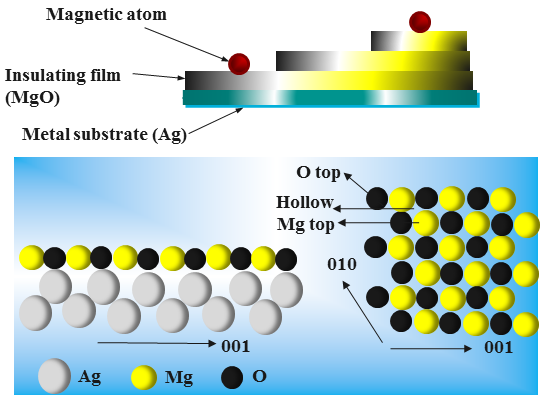}
\caption{Possible binding sites for Mn on the MgO(100) surface.}
\label{fig1}
\end{figure}

The physical system which we will be studying consists of isolated atoms of Mn on a monolayer (ML) (or bilayer) of MgO on a Ag substrate. The MgO surface has three possible binding sites for the Mn: on the O, on the Mg, or the hollow site, midway between two O and two Mg atoms (see Fig.\ \ref{fig1}) In a separate study, we have shown that the O top site is the most stable, followed closely in energy by the hollow site.\cite{hossein} The Mg site is higher in energy and less likely. We find the markedly different charge and spin properties on the three sites. After explaining our calculational method, we present our results for the three sites, comparing and contrasting the calculated spin, charge, and densities of states. 

A previous DFT study has been performed \cite{illasprb} on an Mn adatom on MgO layers, but in the absence of a metallic substrate. A floating layer of MgO appears to favor the O top site for the Mn exclusively, but the addition of Ag layers changes the picture. Each \textit{3d} transition metal atom on the MgO/Ag (001) surface has its signature behavior, such as Co showing maximum anisotropy,\cite{science2014} Ni showing no spin \cite{oliver} and here Mn showing no steps in the inelastic tunneling spectrum (IETS). Using DFT, we explain why no steps were observed even though Mn possesses a high spin configuration.

\noindent \textbf{Results and Discussion}\\
During geometrical coordinate optimization while keeping crystal cell parameter unchanged, we observed the MgO surface rippled at the nearest neighbors (nn) to the adatom. The structural behavior of the surface is very different from site to site. For the O-top and Mg-top cases, this rippling effect did not propagate beyond next nearest neighbor (nnn), but the hollow site was an exception, necessitating us to increase the dimension of supercell from $3\times3$ to $4\times4$ to remove the effect of exchange and structural interactions from neighboring lattice sites. We progressively increased the size of the supercell until all interactions of the adatom region with neighboring cells were null. Due to strong correlation effects of metal \textit{d} orbitals GGA alone is not sufficient to model our system when a Mn adatom is added to the MgO(001)/Ag(001) layer. We found we needed to add an on-site Hubbard $U$ at the Mn 3\textit{d} orbitals at the three different adatom sites.

\begin{figure}[ht]
\centering
\includegraphics[width=1.0\linewidth]{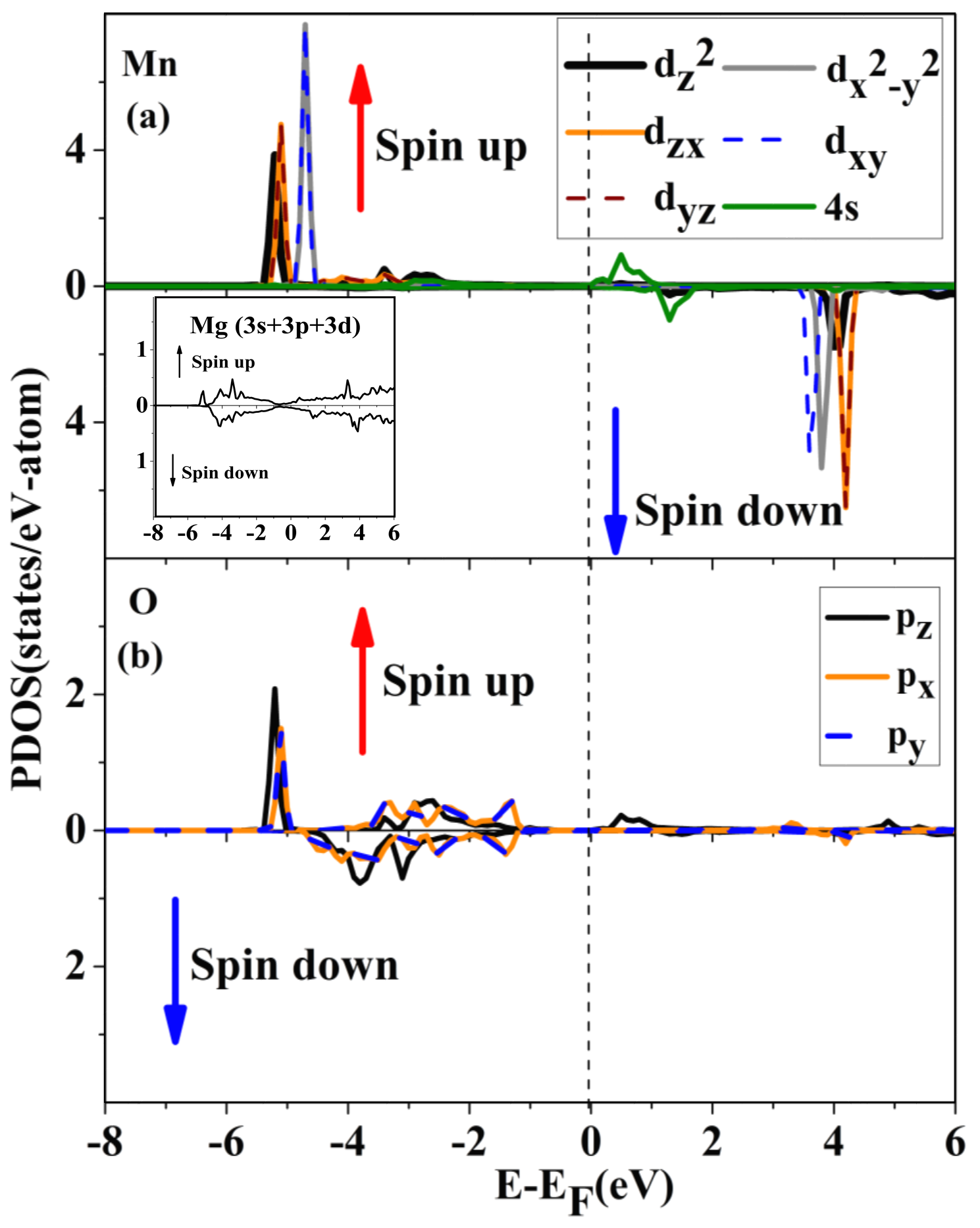}
\caption{Orbital projected density of states (PDOS), showing valence electrons of Mn, O(nn) and Mg(nn) for O top adatom site geometry, using DFT$+U$ ($U=4.0$ eV). (a) Five different 3\textit{d} orbitals and 4\textit{s} orbital of Mn, with PDOS for Mg(nn) plotted in inset. (b) Three \textit{p} orbitals of the O sitting below Mn. In this PDOS plot and also in Fig. \ref{fig5} and Fig. \ref{fig7}, the top half of each plot shows positive spin polarized contribution and the bottom half shows negative spin polarization. The horizontal axis is in eV with energy relative to Fermi level and vertical axis is in states/eV -atom/formula unit. We keep the same axes in the other two PDOS plots (Fig. \ref{fig5} and Fig. \ref{fig7}). Two important facts we note: first, no significant orbital contribution from Mn or O is observed from $-1$ eV to the Fermi level (0 eV). Second, Mn \textit{d} orbitals are split into two degenerate sets, ($d_{z^{2}}$, $d_{zx}$, $d_{yz}$) and ($d_{x^{2}-y^{2}}$, $d_{xy}$), with the first \textit{d} orbital set showing strong overlap with O\textit{ p} orbitals. }
\label{fig2}
\end{figure}

In Table \ref{table1}, we give a brief comparison of calculated Hubbard $U$ (eV), relative energy, selected bond lengths, and magnetic moment ($\mu_{B}$) of Mn among the three adatom sites (atomic Mn has a spin of 2.5 $\mu_B$ \cite{barbara}). The Hubbard $U$, which is a measure of onsite interaction, shows clear variation among the three binding sites. In particular, for a Mn atom at the hollow site, the $U$ is notably higher by 0.6 eV than the values obtained for the O and Mg top sites. The magnetic moment of Mn also charges, with the largest deviation from atomic spin found for the Mg site, though in no case is the atomic magnetic moment of Mn maintained, which suggests that Mn participates to certain amount in covalent bonding at all three possible binding sites. We will discuss the nature of the bonding of each site using projected atomic orbital based densities of states (PDOS) in next few paragraphs.

From Table \ref{table1}, the relative energies of the three configurations show Mn prefers to sit on top of the O atom, whereas the Mg top site is the least preferred. The hollow site and O top site have strong competition, and it is not possible to predict the stable-most binding site unless one uses an appropriate $U$ value. The details are discussed in a previous paper.\cite {hossein} The other notable fact is that the distance from Mn to the next neighboring O atom is roughly the same, within 15\%, for the three adatom positions, which suggests that the Mn$-$O (nm) bond distance is a leading factor which decides the geometries of the three adatom sites. Next, we will describe the nature of the electronic and magnetic structure and bonding of the three different adatom sites. 
\begin{table*}
\begin{tabular}{c c c c}
\hline
 &\begin{tabular}{c} Mg top site \end{tabular} &\begin{tabular}{c} O top site \end{tabular}&\begin{tabular}{c} Hollow site \end{tabular}  \\                                                       
\hline                                                     
\begin{tabular}{c} Hubbard $U$ (eV)  \end{tabular}         &3.8           &4.00               &4.60 \\
Mn$-$Mg (\AA)    & 2.42          &2.07               &2.51\\

Mn$-$O (\AA)     & 1.98          &1.99               &1.71\\

Relative Energy (eV)     & 2.04          &0.00              &0.3 \\

Mn magnetic moment ($\mu_B$)    & 2.2           &2.4               &2.4 \\

\hline
\end{tabular}
\caption{Hubbard $U$, selected bond lengths, relative energy, and Mn magnetic moment with respect to the three different adatom sites shown in Fig. \ref{fig1}. The number of unpaired electrons is twice the spin.}
\label{table1}
\end{table*}

\noindent \textbf{Mn on O top site}\\
The geometry of the O top site shows that Mn is positioned two \AA\ above the O. The MgO(001) surface shows minuscule deformation after geometrical relaxation; the MgO(001)/Ag(001) distance is on average 2.62 \AA\ for Mn-MgO(001)
-Ag(001), as compared to 2.65 \AA\ for the bare system (MgO-Ag). Our calculations yield a Mn 3\textit{d} orbital magnetic moment of 2.4 $\mu_B$, which is very close to the atomic value of 2.5 $\mu_B$. The L\"owdin charges on the Mg(nn) atoms are very low, pointing to a predominant Mn$-$O bonding. The O atom gains about 0.5 electrons.

The PDOS of Mn \textit{d} and \textit{s} valence electrons and O(nn) \textit{p} electrons are shown in Fig. \ref{fig2}. We can compare this result to recently published work \cite{science2014} for a Co adatom on the same surface and O top site. In both cases bond lengths and other geometrical factors are almost the same, but electronic structure and magnetic properties exhibit striking differences. Fig.\ \ref{fig2}(a) shows the five occupied \textit{d} orbitals of Mn, which are nearly degenerate, the crystal field (CF) splitting energies between them never exceeding 1 eV. According to the site symmetry which is related to CF splitting of Mn of O atom top site, Mn \textit{d} orbitals can be divided into three sets as observed in Fig. \ref{fig2}(a): the first one is ($d_{z^{2}}$), the next one is ($d_{zx}$,$d_{yz}$), and the last one is ($d_{x^{2}-y^{2}}$, $d_{xy}$).

But unlike Co, where the ($d_{x^{2}-y^{2}}$, $d_{xy}$) orbitals  are separated from $d_{zx}$  and $d_{yz}$  by 5 eV, the energy separation value for Mn is only 0.05 eV. We can attribute this fact to the increase in CF splitting going towards a higher atomic number along a row in the periodic table. Hence, going from Mn towards Co causes larger CF energy. This lesser CF and large degeneracy means that Mn on O top behaves more like an isolated atom, in contrast to the strongly hybridized Co on O top site of MgO.\cite{science2014} We can also note that the occupied \textit{d} orbitals of Co spread over a range of 7 eV, but for Mn the spread is only 1 eV. 

In addition, for Co on O top we observed a gap of 4.5 eV between highest occupied and lowest unoccupied \textit{d} orbitals, whereas for Mn, the gap is almost twice as large. The PDOS plot in Fig. \ref{fig2} clearly shows a strong Mn $d_{z^{2}}$ and O $p_{z}$ overlap. The $p_{x}$ and $p_{y}$ from O(nn) have peaks at the same energy (Fig. \ref{fig2}(b)) as peaks in the degenerate $d_{zx}$ and $d_{yz}$ of Mn, indicating involvement of the O atom beneath. But we can see from Fig. \ref{fig2}(a) that $d_{xy}$ and $d_{x^{2}-y^{2}}$ have almost double the number of states over $d_{z^{2}}$, $d_{zx}$ and $d_{zy}$, indicating the latter three orbitals participated in hybridization more than the first two. In-plane Mg atoms participate very little in bonding to the Mn. Analysis of the PDOS (Fig. \ref{fig2}) also shows there is no contribution from any Mn \textit{d} and O \textit{p} orbitals near the Fermi level, with just a little \textit{4s} contribution. Those \textit{4s} levels yield a spin polarization of nearly zero because negative and positive spin polarization are both present in equal amounts. We later explain how this has direct consequences regarding the STM experiment.
 
\begin{figure}[ht]
\centering
\includegraphics[width=0.9\linewidth]{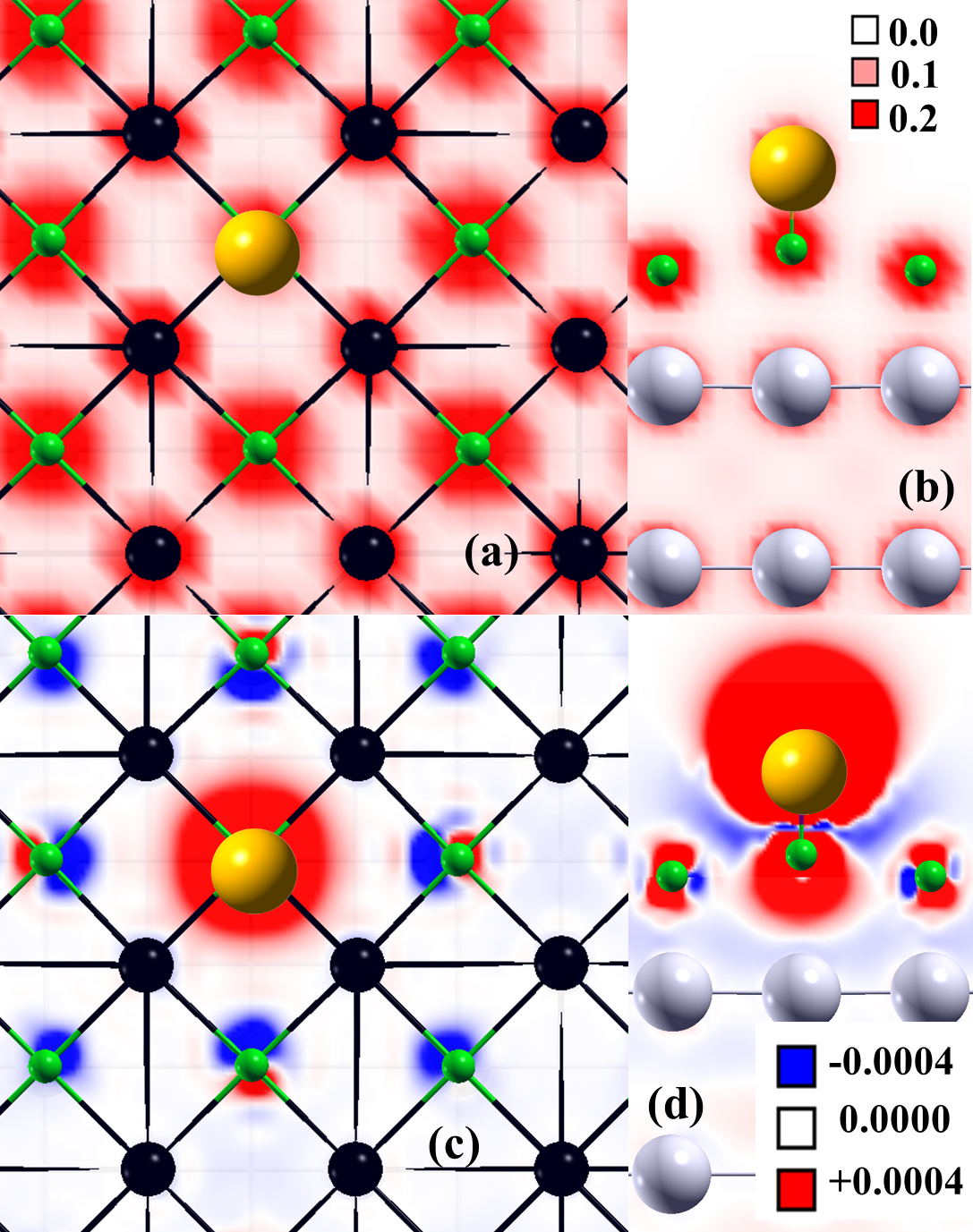}
\caption{Top and side views of charge (panel a and b) and spin density (panel c and d) isocontours\cite{xcrysden} for Mn at O top site using DFT+U (U=4.0 eV). In this figure and also in Fig. \ref{fig8}, panel (a) and (c) shows a projection along the xy plane, whereas (b) and (d) are side projections showing Ag layers and MgO side views. Here and as in Fig. \ref{fig6} and Fig. \ref{fig8}, yellow balls represent Mn, light green balls are O, Mg is colored as black and remaining silver color balls are Ag. Both panels (c) and (d) (also in Fig. \ref{fig8}) reflect red as positive spin and blue as negative spin polarization. From panel (c) and (d) we can see Mn and O are coupled ferromagnetically.}
\label{fig3}
\end{figure}  
 
 The charge distribution of Mn O top site is illustrated in Figs.\ \ref{fig3}(a-b) and the spin density is shown in Fig. \ref{fig3}(c) and (d). Two different projections of geometry are  used, one a top view showing MgO(001) plane with Mn, another a side view showing a vertical view of Ag/MgO/Mn. We followed the same approach of projection for the other two adatom sites also. Charge density from Fig. \ref{fig3}(a) shows different geometrical charge patterns for Mg and O. O charge density is circular, whereas, for Mg, the charge density is squared off. Fig. \ref{fig3}(b) shows a slice of O-Mn with Ag below, and here we can see the anomalous charge distribution does not propagate to the Ag layer, with the MgO showing a somewhat insulating effect.

 Now coming to spin density contours (Figs.\ \ref{fig3}(c-d), we can see that Mn is the most prominent magnetization carrier and no spin density is shown around Mg. Mn is positively spin-polarized and induces ferromagnetic spin polarization in both the O atoms beneath it and those at the next nearest neighbors. From the side view Fig. \ref{fig3}(d) of spin density, one can observe that the spin polarization is not uniform, but rather there is a region of negative spin density in between the two ferromagnetic atoms, because of the local symmetries which require nodes in the interface between the Mn and O, creating a region of partial spin screening. We see this cloud of negative spin polarization for all three sites. The spin polarization of O(nnn) dies down when a Mg atom sits between Mn and O. But from Fig. \ref{fig3}(c) we can see, interestingly, when no Mg is present between Mn and O, O(nnn) on MgO surfaces are ferromagnetically coupled with Mn. From Fig. \ref{fig3}(d), we see a hugely positive spin cloud around Mn, which is circular. This spherical spin cloud of Mn is coming from the nearly degenerate \textit{d} orbitals, discussed above.

Similarly, the spin cloud at the O beneath the Mn in Fig. \ref{fig3}(c) and (d) looks spherical, because $p_x$, $p_y$ and $p_z$ are nearly degenerate, but O(nnn) shows a more \textit{p}-orbital-shaped spin density, in Fig. \ref{fig3}(d). The Mn-induced magnetization of the O beneath it spreads down to nearly the Ag layer. Both charge density and spin density plots indicate strong Mn$-$O $\sigma$-type bonding, which is localized to that site. The non-spin-polarized PDOS of Mg is reflected in the lack of spin polarization seen in Fig. \ref{fig3} at the Mg sites. Though we observe some truncation of Mg electrons to \textit{3p} and \textit{3d} orbitals, bonding of Mg and Mn for O top site is negligible.

\begin{figure}
\centering
\includegraphics[width=0.9\linewidth]{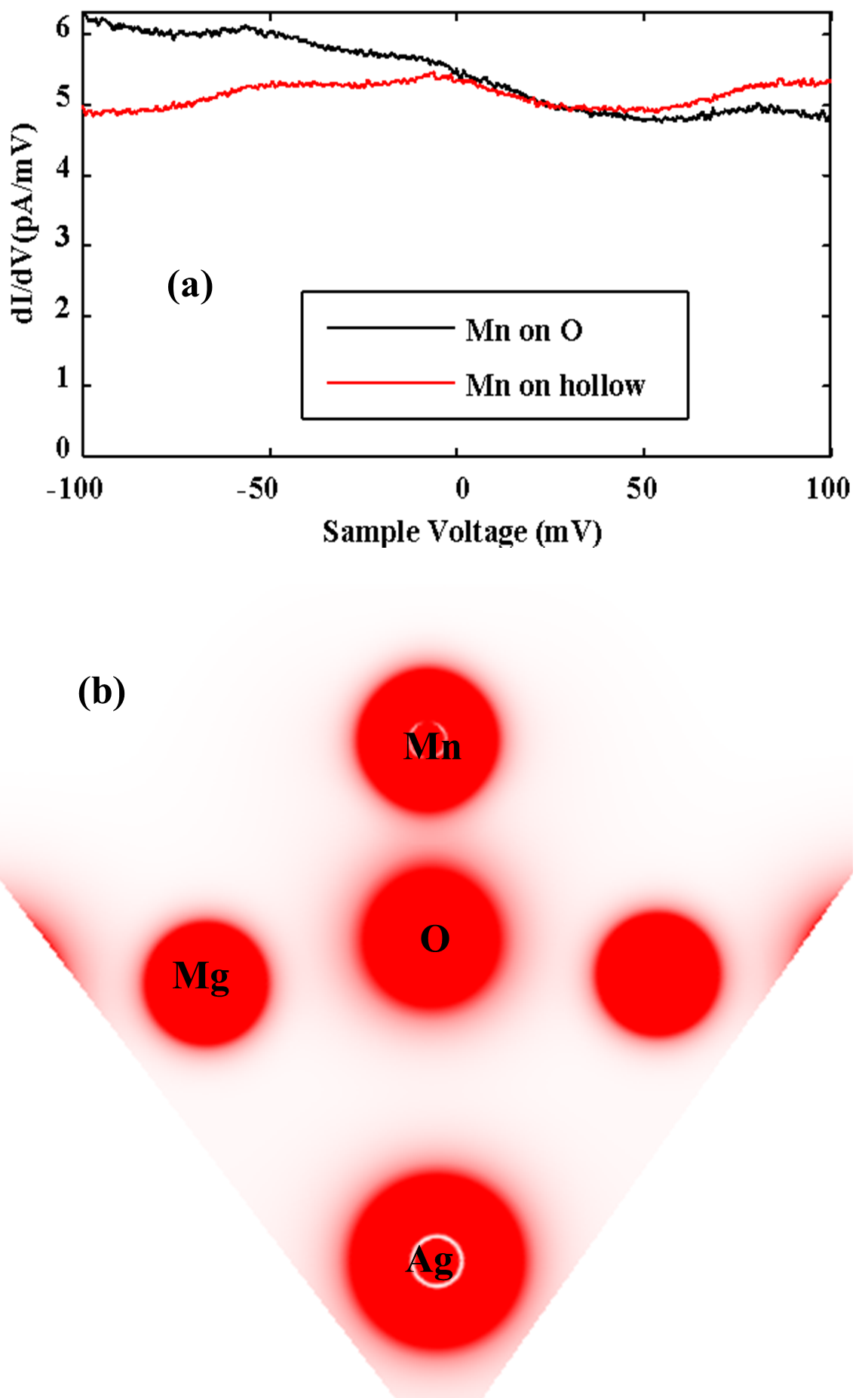}
\caption{(a) IETS spectrum of Mn on both O top and hollow sites. No IETS step observed (b) Charge density plot for Mn on O top with an energy range of $\pm$ 0.1 eV with respect to the Fermi level, showing spherical charge contours for Mn, Mg, O, and Ag}
\label{fig4}
\end{figure}

\noindent \textbf{Experimental Details: Scanning Tunneling Microscopy}\\
Mn, Fe, and Co atoms were deposited on a cold ($\sim$5-10 K) one ML MgO film grown epitaxially on a Ag(001) crystal, as described previously.\cite{science2014} Inelastic electron tunneling spectroscopy (IETS)\cite{stm19,stm29,stm30,stm31,stm32} are applied to probe the quantum spin states of the Mn atoms.\cite{science2014} STM measurements were performed the IBM ultra-high-vacuum low-temperature system.\cite{stm29} IETS measurements are performed by applying a DC voltage between the STM tip (positioned over the Mn atom) and the sample and measuring the conductance using a lock-in technique with a 150 $\mu$V, 806 Hz AC excitation. For inelastic excitations, it is generally observed that when the applied DC voltage is below a threshold of excitation, the conductance is constant, but a sudden increase in conductance is observed when the applied voltage is increased above this excitation energy.\cite{stm29} In such a measurement, electrons tunneling from the STM tip may transfer energy and angular momentum to a magnetic atom and induce spin-flip excitations above a threshold voltage. Fig. \ref{fig4}(a) shows the IETS spectrum of a Mn atom on one ML MgO at 1.2 K. We have not seen any IETS step for Mn up to 7 T magnetic fields. This is a spectrum taken with the same tip over the same atom before/after hopping.

To investigate a possible reason for no IETS steps observed for the Mn/MgO system, we looked at an ``energy range specified” charge density plot. Since IETS is mostly focused on approximately $\pm 0.1$ eV around the Fermi level, we choose that region.  Since O top is the preferred and therefore most likely site proved both by DFT$+U$ and STM, we perform an isocontour plot for only O top geometry here. From  Fig. \ref{fig4}(b), we can see for  O top geometry, the charge is not localized on Mn predominantly; instead, charge density is localized on Mg and O on the surface, and Ag beneath also shows significant charge density. All the charge density contours are circular (spherical or hemispherical), implying a lack of any directional preference, and matching the images seen by STM.  We cannot negate the possibility of low lying excited states where the spin-flip is possible but likely that the energy range is beyond that used by the STM experiment. 

\begin{figure}
\centering
\includegraphics[width=1.0\linewidth]{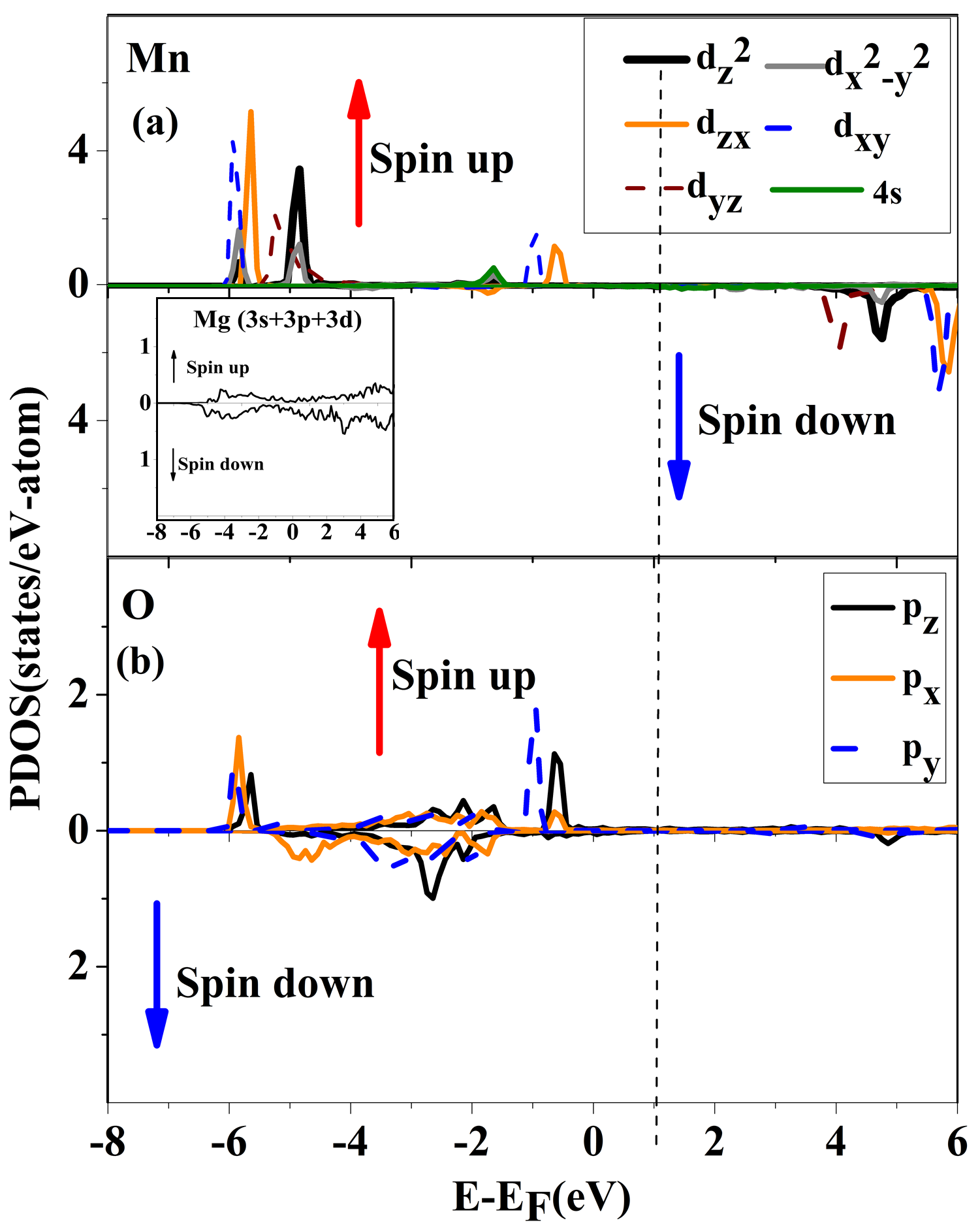}
\caption{PDOS showing valence electrons of Mn, O(nn) and Mg(nn) for hollow site geometry, using DFT$+U$ ($U=4.6$ eV). (a) A five different \textit{3d} orbitals and \textit{4s} orbitals of Mn, with PDOS for Mg(nn) plotted in inset. (b) Three \textit{p} orbitals of the O(nn) coordinated with Mn. We kept the same axis format, as in Fig. \ref{fig2}. Here Mn \textit{d} orbitals are not divided in degenerate sets as in Fig. \ref{fig2}, instead in plane $d_{x^{2}-y^{2}}$ is split into two peaks below the Fermi level.}
\label{fig5}
\end{figure}

\noindent \textbf{Mn on the hollow site}\\
When Mn is sitting on the hollow site, it is coordinated with two Mg and two O atoms. The strong Mn-O bonds keep the O atoms roughly in the same plane. The O atoms rise slightly; conversely, the Mg’s are pushed down towards the Ag layer, after structural relaxation. For this particular adatom configuration, we used a $4 \times 4$ unit cell, because $3\times3$ MgO supercell is not large enough to study single atom magnetism, as significant spin density propagates to the nearest neighbor O. Comparing Fig. \ref{fig2}(a) and Fig. \ref{fig5}, we can clarify this phenomenon: occupied \textit{d} orbitals of Mn are spread over a wider region (6 eV) for the hollow site Fig. \ref{fig5}(a), compared to O top (1 eV) Fig. \ref{fig2}(a). This indicates a very nonlocalized bonding present in the hollow site, whereas for O top site, the effect of the magnetic atom is very localized. Mn$-$O bond lengths are 1.71 \AA\ and Mn$-$Mg bond lengths are 2.51 \AA. The distance between O/Mg and the first Ag(001) layer is found to be 2.73/2.66 \AA. We obtained a Hubbard $U$ value of 4.6 eV from first-principles calculations using the linear response approach implemented in Quantum Espresso.\cite{U_qe}Using PBE$+U$ (U=4.6 eV), a magnetic moment of 2.2 $\mu_B$ is obtained for the Mn atom, which points to a more covalent interaction between O atoms than the O top case. Though here, instead of one O, Mn is interacting with two oxygens in the MgO (001)lattice.

 \begin{figure*}
\includegraphics[width=0.9\textwidth]{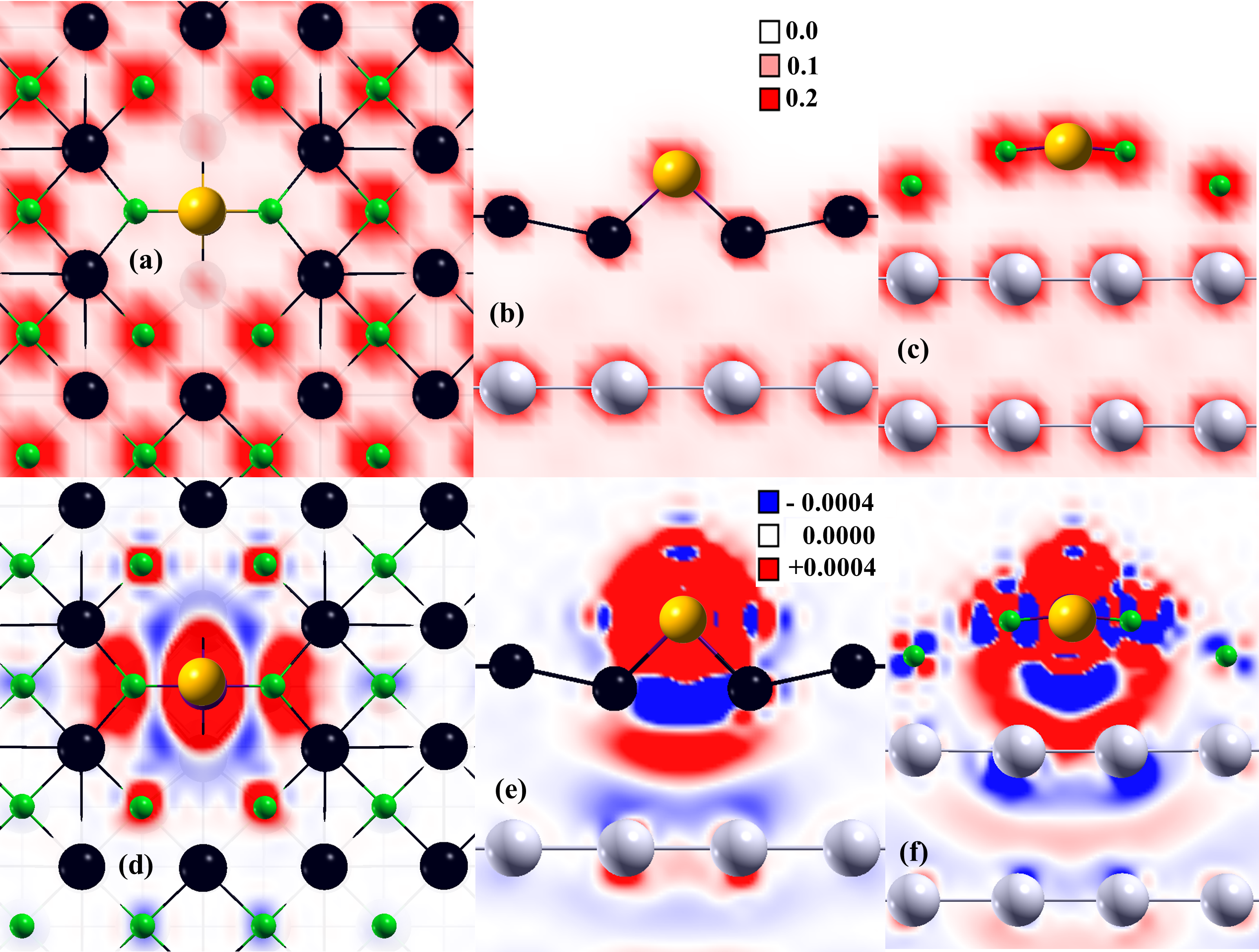}
\caption{Charge density (a)-(c) and spin density (d)-(f) isocontours of Mn adatom on hollow site for $4 \times 4$ supercell obtained using DFT$+U$($U=4.6$ eV). (a) and (d) charge and spin density in xy plane. (b) and (e) charge and spin density showing Mg-Mn bonding, (c ) and (f)   plotted same as for O-Mn bond. We use same color index as in Fig. \ref{fig3}. Mg(nn) is pushed down closer to the Ag layer, but O(nn) is closer to Mn. O(nn) and Mn are ferromagnetically coupled, same as O otop site. Mn spin density (c) shows elliptical shape, in contrast to circular spin density we have for O top site.}
\label{fig6}
\end{figure*}
 
From the PDOS plot Fig. \ref{fig5}, we can see the degeneracy of Mn \textit{d} orbitals is somewhat destroyed. Now instead of along the Z axis, Mn and O are aligned on the xy plane.  This fact can be clearly seen in the  PDOS plot Fig. \ref{fig5}: two in plane \textit{d} orbitals of Mn, $d_{x^{2}-y^{2}}$, $d_{xy}$, are degenerate at around -5.8 eV. In this energy region $d_{x^{2}-y^{2}}$, $d_{xy}$ show overlap between O $p_{x}$, $p_{y}$ and $p_{z}$. $d_{x^{2}-y^{2}}$ is participating in $\sigma$ bonding but $d_{xy}$ shows $\pi$ overlap too. Now this $\sigma$ interaction is the reason behind the coplaner nature of the Mn-O bond. $d_{x^{2}-y^{2}}$ and $d_{z^{2}}$ are split into another peak at $-4.8$ eV, where for O, an opposite spin polarized O $p_{x}$ is present. This overlap leads to an antibonding interaction, making the hollow site less stable than O top. If we go back to Fig. \ref{fig2}, it is clear that there is no antibonding overlap present between Mn(d) and O (p). Fig. \ref{fig5}(a) has another difference from Fig. \ref{fig2}(a), which is, for O top, none of the occupied \textit{d} orbitals split into two peaks, whereas for hollow site, $d_{x^{2}-y^{2}}$, $d_{xy}$, and $d_{zx}$ split into two asymmetrical peaks. This is clearly due to the less symmetric nature of the hollow geometry.

The $d_{xy}$/ $d_{zx}$ split as twins, the first peaks appearing at $-6$ eV/$-5.5$ eV below the Fermi level, and the second twin peaks appearing at $-1$ eV/$-0.5$ eV. In both cases, we can see a bonding overlap between Mn and O. The differences between high and low energy peaks is, close to the Fermi level, Mn and O show almost equal contribution, but at -6 eV, the intensity of the Mn d peaks is three times larger than the coordinated O atom. For both sets of peaks, there are $\pi$  overlaps, $d_{xy}$ with O $p_{y}$ and $d_{zx}$ with $p_z$. $d_{z^{2}}$ remains non bonded because of the absence of z axial O, unlike O top geometry.

Now considering charge and spin density, Fig. \ref{fig6}, all six subsections of this figure indicate charge and spin contours are no longer as symmetric as O top site (Fig. \ref{fig3}). Here O shows a square-shaped charge density due to $p_x$ and $p_y$ degeneracy. Fig. \ref{fig6}(a) (b) and (c) are plotted to show the charge distribution along three different projections, \textit{xy}, \textit{xz} and \textit{yz} planes. Fig. \ref{fig6}(d) (e) and (f) are spin density isocontours using the same projections. The charge contours show Mn bonding interaction in the \textit{xy} plane with O in Fig. \ref{fig6}(c). Spin density contours indicate that the Mn atom again polarizes the next-nearest neighbor O atoms. The spin minority cloud around the Mn atom extends towards the next-nearest neighbor O atoms in the xy plane, much broader than the spin minority cloud around the Mn when in O-top position.

Interestingly, in the xy plane, we find spatial regions of alternating spin with the periodicity of the underlying Ag row. Even the topmost layer of Ag atoms becomes spin-polarized, which shows the spin leaking beneath the insulator, see Fig. \ref{fig6}(e) and (f). The Mn spin density has an elliptical shape along the \textit{xy} plane. Fig. \ref{fig6}(f) shows both antibonding and bonding interactions between Mn and O, as we inferred from the PDOS plot in Fig. \ref{fig5}. There is a negative spin cloud present between MgO and Ag layer, which is not present for O top. The reason for that spin cloud is, with Mn sitting in a hollow site, it is $d_{z^{2}}$ is almost  non hybridized with O \textit{p}orbitals, so $d_{z^{2}}$ penetrates to the gap between MgO and Ag, causing significant spin leaking through MgO and turning the interface Ag layer spin-polarized. Though this adatom site is highly stable, the presence of an in-plane antibonding interaction lowers the stability of hollow site a bit over O top site.

\noindent \textbf{Mn on Mg top site}\\
The bonding of transition metal atoms with the Mg in MgO is important for hydrogen storage \cite{Zhou}. Unlike the other two configurations in which the Mn adatom stays well separated from the MgO layer, with Mn on top of it, the Mg atom unexpectedly sinks towards the Ag(001) layer below. We see a possible source for this odd move by referring to the consistency of the Mn-O bond length noted previously (Table 1). It seems the Mn-O bond is so strong that even on top of the Mg atom, the Mn atom adjusts its position so that it achieves the preferred distance of roughly 1.9 \AA\ to O. To enable this Mn-O spacing, the Mg atom has to sink down. From the strong structural relaxation, Mn does not maintain its full magnetic moment, but we find a reduction of 12\%, which may explain why this is the least stable configuration. 

\begin{figure}[t]
\centering
\includegraphics[width=1.0\linewidth]{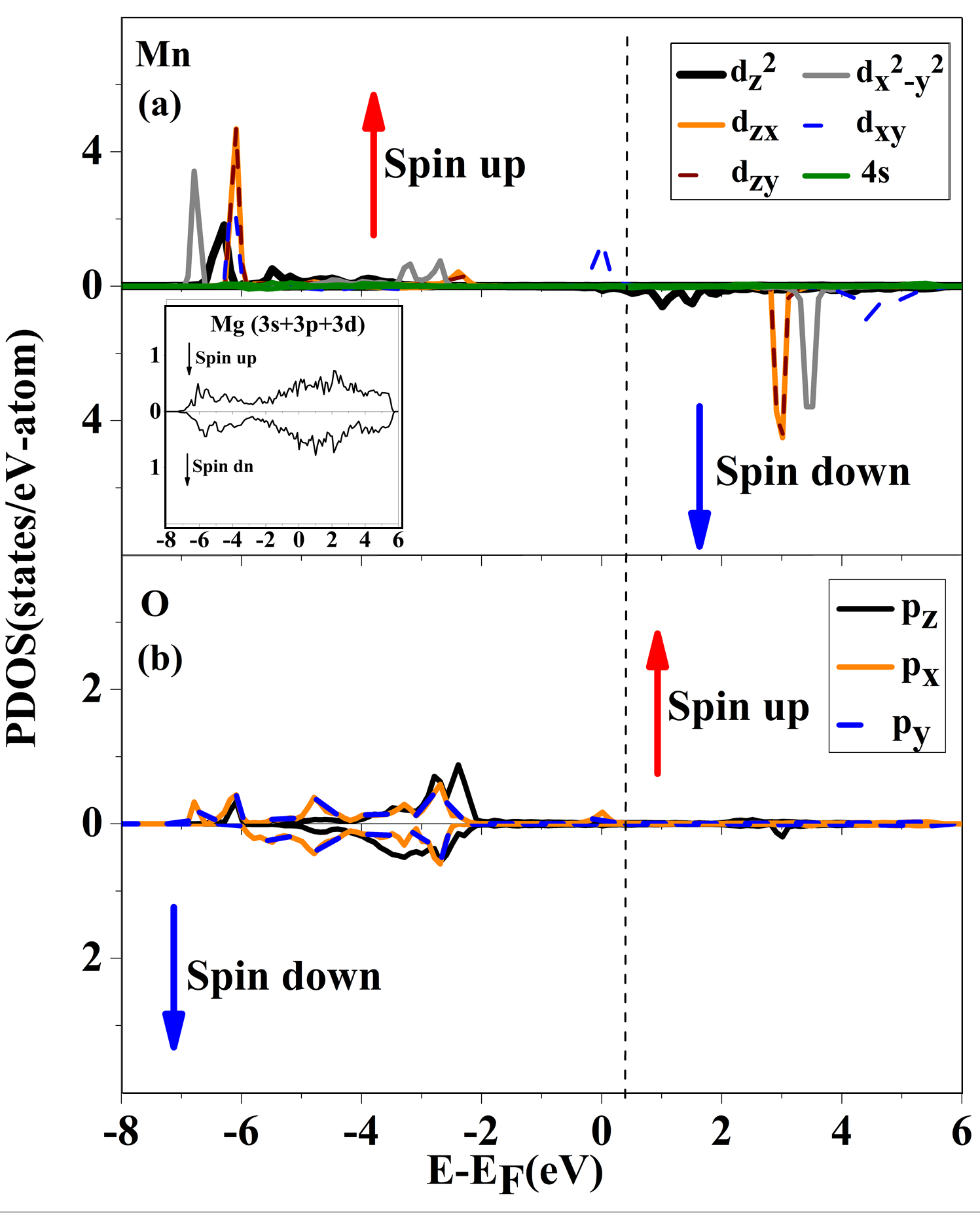}
\caption{PDOS showing valence electrons of Mn, O(nn), and Mg(nn) when Mn sits on Mg top, using DFT$+U$ ($U=3.8$ eV), we used same axis notations and panel indexes as in Fig. \ref{fig2}  and Fig. \ref{fig5}. We can see at the panel (a) nearly two degenerate \textit{d} orbital sets, but not the same as Fig. \ref{fig2}(a), clearly showing the different type of hybridization than O top geometry.}
\label{fig7}
\end{figure}

The PDOS plot for Mn atom at Mg top site Fig. \ref{fig7} is the most complex of the three cases. There are three prominent peaks at low energies, with a very low response from the oxygen. The inset, showing Mg, shows that the Mg, also, has most of its weight closer to the Fermi energy. The Mg, so close to the Ag layer, becomes primarily coupled to Ag. The lack of symmetry of the Mn peaks at -7 eV is due to the coupling to the four O atoms which are its nearest neighbors. 

\begin{figure}[t]
\includegraphics[width=0.5\textwidth]{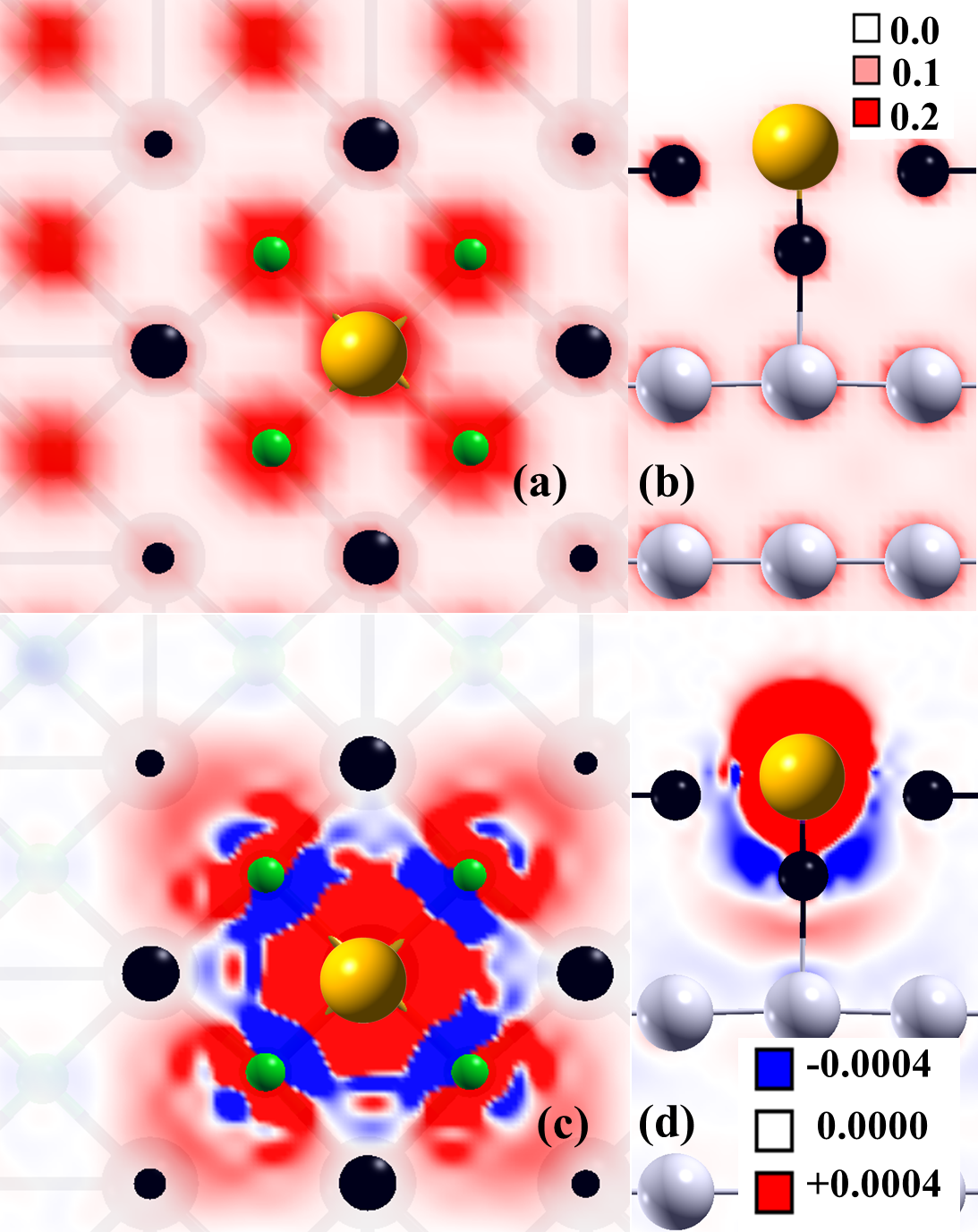}
\caption{ Top and side views of charge (panel a and b) and spin density (panel c and d) isocontours for Mn on Mg top site using DFT$+U$ ($U=3.8$ eV) . Here Mn pushes  Mg far below towards the Ag layer. (a) Mn coordination with four nearest O.  (b) show no bonding overlap between Mn-Mg or Ag-Mg, (c) shows antiferromagnetic cloud between Mn and O(nn), with complex O(nn) spin polarization. }
\label{fig8}
\end{figure}

The calculated charge density and spin density contours for Mg otop geometry are shown in Fig. \ref{fig8}(a) and (d). Fig. \ref{fig8}(a) and (c) are charge, and spin contours of the xy plane, and (b) and (d) are for the xz plane. The Mn at Mg top is a complex structure with four oxygen atoms coordinated with Mn in a coplaner fashion to make a square planer Mn-O$_{4}$ unit. From Fig. \ref{fig8}(a) we can see the O atoms have a circular charge contour, coming from three degenerate \textit{p} orbitals (see PDOS Fig. \ref{fig7}(b)). The square shape charge contour of Mn comes from two separate degenerate \textit{d} orbital sets, one a nonbonded “$d_{x^{2}- y^{2}}$ and $d_{xy}$” combination, another from the contribution from the x component of $d_{xz}$ and y component of $d_{yz}$. Mn $d_{zx}$, $d_{yz}$ and O atoms  $p_x$ and $p_y$ from  combine here along the x and y axis and form a $\pi$ bond.

The nature of the spin contours (Fig. \ref{fig8}(c) and (d)) for this adatom geometry has some unique features. The dramatic sinking of the Mg atom towards the Ag layer after adding Mn is shown in Fig. \ref{fig8}(b)(black atom under the yellow). Although from charge density, there does not seem to be a bonding overlap between Mn and Mg, due to their separation, this geometry gives rise to a special spin density whose contour plots are shown in Fig. \ref{fig8}(c) and (d). From the PDOS (Fig. \ref{fig7}), all participating $\pi$ bonding orbitals are present as spin up, but between Mn and O, we can see the presence of a significant spin-down cloud, reminiscent of the spin-down clouds separating the Mn and O in the other geometries. 

The spin minority cloud here can only come from O atoms because Mn has no occupied spin negative density of states. The surrounding O atoms show a complex spatial distribution of the two spin channels. The net polarization of the O atoms is thus reduced compared to the other two configurations because now Mn is sharing its spin density between four O atoms. Also, we see the spin leaking effect towards the Ag layer is diminished, compared to the hollow site, even though there is a path for spin density to the Ag layer around the sunken Mg. This indicates that to prevent spin leakage from the surface layers, the adatom needs to sit on a surface atom. Overall the Mg top site is energetically less favorable (quasi-stable only locally). But electronic structure calculations reveal some very interesting structure and bonding phenomena, which is useful in understanding atomic-scale magnetism.

\noindent \textbf{Conclusions}\\
We have successfully modeled the electronic and magnetic structure of the Mn/MgO/Ag system using PBE$+U$. Mn has three symmetrically allowed potential binding sites on MgO(001). These sites are on oxygen, in between two Mg and two O (hollow), and on Mg top. Since the bonding chemistry is quite different from site to site, we used first-principles methods to determine Hubbard $U$. The $U$ values varied for the different adatom sites ($U_{O}$=4.0 eV, $U_{Hollow}$=4.6 eV, and  $U_{Mg}$=3.8 eV), and using these $U$ values we find Mn prefers the  O site, with hollow site energetically very close.  Atom projected density of states clearly show that Mn on O top has very low crystal field splitting among its \textit{d} orbitals, and retains nearly its atomic magnetic moment, though Mn \textit{d} orbitals are hybridized to some extent with nearest O \textit{p$_z$} orbitals. 

Mn at the hollow site shows a little less spin moment then O top site, as mainly Mn is now shared between two Mg and two O atoms. In this case, Mn \textit{d} orbital shows larger crystal field splitting compared to O top. In the hollow site, spin propagation on the xy plane is quite dispersed, with the spin density contour showing that  Mn spin density propagates to the second nearest neighbor O atoms. Hollow site geometry also shows interesting spin contours, with O atoms as the nearest and the next nearest neighbor from Mn displaying ferromagnetic exchange with Mn, whereas the second next nearest neighbor O atoms have an antiferromagnetic orientation.

In addition, we provided a first principles-based explanation of why we have not observed any IETS step in the spin $\frac{5}{2}$ Mn atom on either O top or hollow sites. Mg top site is energetically the least favored. In this geometry, we see Mn pushes Mg down beneath it, towards the closest Ag layer. This phenomenon overall results in a very long Mg-Mn distance, with Mn now coordinated with four O atoms and having an Mn-O $\pi$ bonding overlap. The common features observed in these three geometries are, a) Mg participates almost not at all in bonding to Mn. b) There is always a spin minority cloud present between Mn-O. We also observe for O top geometry, spin leaking towards the Ag layers shows the least intensity, with the Mn-O unit keeping most of the spin to itself. There is increasing leaking to the Ag on the Mg site, and most of the hollow site, due to unobstructed access to the Ag layer by the Mn. We conclude that PBE$+U$ is an important tool for understanding transition-metal interfaces with complex surfaces. Future work will include the effects of inter-Mn exchange coupling. These systems have many important applications, including possible magnetic memory \cite{jap} as well as energy storage devices.\cite{nature}  

\noindent \textbf{Methods}\\
To model a system of an adatom on a surface using DFT, we must first transform the whole Mn/MgO/Ag into a periodic system. We used slab geometry, in which an ML of MgO is added to six Ag layers, and this system repeated infinitely, with a vacuum of 15 \AA\ between the slabs. Likewise, on the surface of the slab, a single Mn adatom is modeled by a periodic array of Mn, with separation large enough to avoid inter-Mn interactions. Our unit cell must then be large enough that a) the ML of MgO behaves as if it is on an infinite layer of Ag; b) the lower Ag edge of one slab does not interact with the top Mn/MgO layer of the one under it, and c) the Mn atoms are far enough apart not to interact. Here we define the Mg or O atom nearest to the adatom as a nearest neighbor(nn) and the next one as next nearest neighbor (nnn). We illustrate the geometry of the surface in Fig. \ref{fig1}. To accurately calculate the properties of this system, therefore, we require a large unit cell. 

Specifically, our slab model consists of six Ag(001) layers with one MgO (001) layer on top.\cite{hossein} To avoid interaction with the next slab, a vacuum of about 15 \AA\ is added in the z-direction between the slabs. We applied periodic boundary conditions in all three directions, throughout our calculations. In Fig. \ref{fig1}. we show the three possible adatom sites at the 001 planes of MgO, namely a) on top of O, b) in between two Mg and two O and c) on top of Mg, hereafter referred to as O top, hollow and Mg top respectively. During relaxation, we fixed the last three Ag layers at bulk value, as well as fixing the xy positions of the atoms at the boundaries of the unit cell, and allowed the rest of the slab to relax. This process was applied to all three adatom sites. We determined that for O top and Mg top geometry, we need a $3\times3$ supercell, whereas, for the hollow site, we need a $4\times4$ supercell. 

The electronic structure calculations are performed within a generalized gradient approximation (GGA) \cite{pbe} framework, using a Vanderbilt ultrasoft pseudopotential\cite{ultrasoft} within the Quantum ESPRESSO \cite{qe} package. We use Monkhorst-Pack $8\times8\times1$ for O top and Mg top geometry and $6\times6\times1$ k-grids for the hollow site. Charge and spin arrangements are sensitive to the initial configurations. We obtained $U$ for each possible binding site using relaxed geometries from first principles methods using Quantum ESPRESSO \cite{qe, U_qe}. For O top and Mg top, we recalibrate Hubbard $U$ using all electron-based WIEN2K \cite{wien2k, epl} code. Both the codes give us similar $U$ values for O top and Mg top site.

\noindent \textbf{Acknowledgments}\\
We would like to thank Dr. Susanne Baumann, Dr. William Paul, Dr. Ileana Rau, Christopher Lutz, Dr. Andreas Heinrich, for their IETS spectrum. Authors gratefully acknowledge suggestions and discussions with Professor Warren Pickett. BAJ thanks the Aspen Center for Physics and the NSF Grant $\#$1066293 for hospitality during the research and writing phases of this work. Our research used resources of the National Energy Research Scientific Computing  Center (NERSC), a DOE Office of Science User Facility supported by the Office of Science of the U.S. Department of Energy under Contract No. DE-AC02-05CH11231. Blue gene hours from KAUST supercomputer center and IBM Almaden Blue gene resources. This research is supported by the King Abdullah University of Science and Technology (KAUST). Authors thankfully acknowledge special support from KAUST high=performance computing team. SG acknowledges DOE Stockpile Stewardship Academic Alliance Program under Grant $\#$ DE-FG03-03NA00071. 

\noindent \textbf{Contributions}\\
S.G performed calculations. S.G., T.P.K., U.S, and B.A.J wrote the manuscript. \\ 
\noindent \textbf{Competing interests}\\
The authors declare no conflict of interest.

\bibliography{mn2phys}

\begin{thebibliography}{44}%
\makeatletter
\providecommand \@ifxundefined [1]{%
 \@ifx{#1\undefined}
}%
\providecommand \@ifnum [1]{%
 \ifnum #1\expandafter \@firstoftwo
 \else \expandafter \@secondoftwo
 \fi
}%
\providecommand \@ifx [1]{%
 \ifx #1\expandafter \@firstoftwo
 \else \expandafter \@secondoftwo
 \fi
}%
\providecommand \natexlab [1]{#1}%
\providecommand \enquote  [1]{``#1''}%
\providecommand \bibnamefont  [1]{#1}%
\providecommand \bibfnamefont [1]{#1}%
\providecommand \citenamefont [1]{#1}%
\providecommand \href@noop [0]{\@secondoftwo}%
\providecommand \href [0]{\begingroup \@sanitize@url \@href}%
\providecommand \@href[1]{\@@startlink{#1}\@@href}%
\providecommand \@@href[1]{\endgroup#1\@@endlink}%
\providecommand \@sanitize@url [0]{\catcode `\\12\catcode `\$12\catcode
  `\&12\catcode `\#12\catcode `\^12\catcode `\_12\catcode `\%12\relax}%
\providecommand \@@startlink[1]{}%
\providecommand \@@endlink[0]{}%
\providecommand \url  [0]{\begingroup\@sanitize@url \@url }%
\providecommand \@url [1]{\endgroup\@href {#1}{\urlprefix }}%
\providecommand \urlprefix  [0]{URL }%
\providecommand \Eprint [0]{\href }%
\providecommand \doibase [0]{http://dx.doi.org/}%
\providecommand \selectlanguage [0]{\@gobble}%
\providecommand \bibinfo  [0]{\@secondoftwo}%
\providecommand \bibfield  [0]{\@secondoftwo}%
\providecommand \translation [1]{[#1]}%
\providecommand \BibitemOpen [0]{}%
\providecommand \bibitemStop [0]{}%
\providecommand \bibitemNoStop [0]{.\EOS\space}%
\providecommand \EOS [0]{\spacefactor3000\relax}%
\providecommand \BibitemShut  [1]{\csname bibitem#1\endcsname}%
\let\auto@bib@innerbib\@empty
\bibitem [{\citenamefont {Freund}\ and\ \citenamefont
  {Pacchioni}(2008)}]{Freund}%
  \BibitemOpen
  \bibfield  {author} {\bibinfo {author} {\bibfnamefont {H.-J.}\ \bibnamefont
  {Freund}}\ and\ \bibinfo {author} {\bibfnamefont {G.}~\bibnamefont
  {Pacchioni}},\ }\href@noop {} {\bibfield  {journal} {\bibinfo  {journal}
  {Chem. Soc. Rev}\ }\textbf {\bibinfo {volume} {37}},\ \bibinfo {pages} {2224}
  (\bibinfo {year} {2008})}\BibitemShut {NoStop}%
\bibitem [{\citenamefont {Robertson}(2009)}]{Robertson}%
  \BibitemOpen
  \bibfield  {author} {\bibinfo {author} {\bibfnamefont {J.~J. B. o.~a.}\
  \bibnamefont {Robertson}},\ }\href@noop {} {\bibfield  {journal} {\bibinfo
  {journal} {J. Vac. Sci. Technol. B}\ }\textbf {\bibinfo {volume} {27}},\
  \bibinfo {pages} {277} (\bibinfo {year} {2009})}\BibitemShut {NoStop}%
\bibitem [{\citenamefont {Netzer}\ \emph {et~al.}(2010)\citenamefont {Netzer},
  \citenamefont {Allegretti},\ and\ \citenamefont {Surnev}}]{Netzer}%
  \BibitemOpen
  \bibfield  {author} {\bibinfo {author} {\bibfnamefont {F.~P.}\ \bibnamefont
  {Netzer}}, \bibinfo {author} {\bibfnamefont {F.}~\bibnamefont {Allegretti}},
  \ and\ \bibinfo {author} {\bibfnamefont {S.}~\bibnamefont {Surnev}},\
  }\href@noop {} {\bibfield  {journal} {\bibinfo  {journal} {J. Vac. Sci.
  Technol. B}\ }\textbf {\bibinfo {volume} {28}},\ \bibinfo {pages} {1}
  (\bibinfo {year} {2010})}\BibitemShut {NoStop}%
\bibitem [{\citenamefont {Greiner}\ \emph {et~al.}(2012)\citenamefont
  {Greiner}, \citenamefont {Chai}, \citenamefont {Helander}, \citenamefont
  {Tang},\ and\ \citenamefont {Lu}}]{Greiner}%
  \BibitemOpen
  \bibfield  {author} {\bibinfo {author} {\bibfnamefont {M.~T.}\ \bibnamefont
  {Greiner}}, \bibinfo {author} {\bibfnamefont {L.}~\bibnamefont {Chai}},
  \bibinfo {author} {\bibfnamefont {M.~G.}\ \bibnamefont {Helander}}, \bibinfo
  {author} {\bibfnamefont {W.-M.}\ \bibnamefont {Tang}}, \ and\ \bibinfo
  {author} {\bibfnamefont {Z.-H.}\ \bibnamefont {Lu}},\ }\href@noop {}
  {\bibfield  {journal} {\bibinfo  {journal} {Adv. Funct. Mater}\ }\textbf
  {\bibinfo {volume} {22}},\ \bibinfo {pages} {4557} (\bibinfo {year}
  {2012})}\BibitemShut {NoStop}%
\bibitem [{\citenamefont {Jaouen}\ \emph {et~al.}(2012)\citenamefont {Jaouen},
  \citenamefont {Jezequel}, \citenamefont {Delhaye}, \citenamefont {Lepine},
  \citenamefont {Turban},\ and\ \citenamefont {Schieffer}}]{Jaouen}%
  \BibitemOpen
  \bibfield  {author} {\bibinfo {author} {\bibfnamefont {T.}~\bibnamefont
  {Jaouen}}, \bibinfo {author} {\bibfnamefont {G.}~\bibnamefont {Jezequel}},
  \bibinfo {author} {\bibfnamefont {G.}~\bibnamefont {Delhaye}}, \bibinfo
  {author} {\bibfnamefont {B.}~\bibnamefont {Lepine}}, \bibinfo {author}
  {\bibfnamefont {P.}~\bibnamefont {Turban}}, \ and\ \bibinfo {author}
  {\bibfnamefont {P.}~\bibnamefont {Schieffer}},\ }\href@noop {} {\bibfield
  {journal} {\bibinfo  {journal} {Appl. Phys. Lett}\ }\textbf {\bibinfo
  {volume} {100}},\ \bibinfo {pages} {022103} (\bibinfo {year}
  {2012})}\BibitemShut {NoStop}%
\bibitem [{\citenamefont {Giordano}\ \emph {et~al.}(2006)\citenamefont
  {Giordano}, \citenamefont {Cinquini},\ and\ \citenamefont
  {Pacchioni}}]{Giordano}%
  \BibitemOpen
  \bibfield  {author} {\bibinfo {author} {\bibfnamefont {L.}~\bibnamefont
  {Giordano}}, \bibinfo {author} {\bibfnamefont {F.}~\bibnamefont {Cinquini}},
  \ and\ \bibinfo {author} {\bibfnamefont {G.}~\bibnamefont {Pacchioni}},\
  }\href@noop {} {\bibfield  {journal} {\bibinfo  {journal} {Phys. Rev. B}\
  }\textbf {\bibinfo {volume} {73}},\ \bibinfo {pages} {045414} (\bibinfo
  {year} {2006})}\BibitemShut {NoStop}%
\bibitem [{\citenamefont {Prada}\ \emph {et~al.}(2008)\citenamefont {Prada},
  \citenamefont {Martinez},\ and\ \citenamefont {Pacchioni}}]{Prada}%
  \BibitemOpen
  \bibfield  {author} {\bibinfo {author} {\bibfnamefont {S.}~\bibnamefont
  {Prada}}, \bibinfo {author} {\bibfnamefont {U.}~\bibnamefont {Martinez}}, \
  and\ \bibinfo {author} {\bibfnamefont {G.}~\bibnamefont {Pacchioni}},\
  }\href@noop {} {\bibfield  {journal} {\bibinfo  {journal} {Phys.Rev.B}\
  }\textbf {\bibinfo {volume} {78}},\ \bibinfo {pages} {235423} (\bibinfo
  {year} {2008})}\BibitemShut {NoStop}%
\bibitem [{\citenamefont {Brongersma}(2003)}]{Brongersma}%
  \BibitemOpen
  \bibfield  {author} {\bibinfo {author} {\bibfnamefont {M.~L. N. p.~N.}\
  \bibnamefont {Brongersma}},\ }\href@noop {} {\bibfield  {journal} {\bibinfo
  {journal} {Nat. Mater}\ }\textbf {\bibinfo {volume} {2}},\ \bibinfo {pages}
  {296} (\bibinfo {year} {2003})}\BibitemShut {NoStop}%
\bibitem [{\citenamefont {Chen}\ and\ \citenamefont {Goodman}(2004)}]{Chen}%
  \BibitemOpen
  \bibfield  {author} {\bibinfo {author} {\bibfnamefont {M.~S.}\ \bibnamefont
  {Chen}}\ and\ \bibinfo {author} {\bibfnamefont {D.~W.}\ \bibnamefont
  {Goodman}},\ }\href@noop {} {\bibfield  {journal} {\bibinfo  {journal}
  {Science}\ }\textbf {\bibinfo {volume} {306}},\ \bibinfo {pages} {252}
  (\bibinfo {year} {2004})}\BibitemShut {NoStop}%
\bibitem [{\citenamefont {Glaspell}\ \emph {et~al.}(2008)\citenamefont
  {Glaspell}, \citenamefont {Hassan}, \citenamefont {Elzatahry}, \citenamefont
  {Abdalsayed},\ and\ \citenamefont {El-Shall}}]{Glaspell}%
  \BibitemOpen
  \bibfield  {author} {\bibinfo {author} {\bibfnamefont {G.}~\bibnamefont
  {Glaspell}}, \bibinfo {author} {\bibfnamefont {M.~A.~H.}\ \bibnamefont
  {Hassan}}, \bibinfo {author} {\bibfnamefont {A.}~\bibnamefont {Elzatahry}},
  \bibinfo {author} {\bibfnamefont {V.}~\bibnamefont {Abdalsayed}}, \ and\
  \bibinfo {author} {\bibfnamefont {M.~S.}\ \bibnamefont {El-Shall}},\
  }\href@noop {} {\bibfield  {journal} {\bibinfo  {journal} {Top. Catal}\
  }\textbf {\bibinfo {volume} {47}},\ \bibinfo {pages} {22} (\bibinfo {year}
  {2008})}\BibitemShut {NoStop}%
\bibitem [{\citenamefont {Luo}\ \emph {et~al.}(2008)\citenamefont {Luo},
  \citenamefont {Lin}, \citenamefont {Wen},\ and\ \citenamefont {Chang}}]{Luo}%
  \BibitemOpen
  \bibfield  {author} {\bibinfo {author} {\bibfnamefont {M.~F.}\ \bibnamefont
  {Luo}}, \bibinfo {author} {\bibfnamefont {W.~R.}\ \bibnamefont {Lin}},
  \bibinfo {author} {\bibfnamefont {W.~H.}\ \bibnamefont {Wen}}, \ and\
  \bibinfo {author} {\bibfnamefont {W.~B.}\ \bibnamefont {Chang}},\ }\href@noop
  {} {\bibfield  {journal} {\bibinfo  {journal} {Surf. Sci}\ }\textbf {\bibinfo
  {volume} {602}},\ \bibinfo {pages} {3258} (\bibinfo {year}
  {2008})}\BibitemShut {NoStop}%
\bibitem [{\citenamefont {Harding}\ \emph {et~al.}(2009)\citenamefont
  {Harding}, \citenamefont {Habibpour}, \citenamefont {Kunz}, \citenamefont
  {Farnbacher},\ and\ \citenamefont {S.}}]{Harding}%
  \BibitemOpen
  \bibfield  {author} {\bibinfo {author} {\bibfnamefont {C.}~\bibnamefont
  {Harding}}, \bibinfo {author} {\bibfnamefont {V.}~\bibnamefont {Habibpour}},
  \bibinfo {author} {\bibfnamefont {S.}~\bibnamefont {Kunz}}, \bibinfo {author}
  {\bibfnamefont {H.}~\bibnamefont {Farnbacher}}, \ and\ \bibinfo {author}
  {\bibfnamefont {A.~N.}\ \bibnamefont {S.}},\ }\href@noop {} {\bibfield
  {journal} {\bibinfo  {journal} {J. Am. Chem. Soc}\ }\textbf {\bibinfo
  {volume} {131}},\ \bibinfo {pages} {538} (\bibinfo {year}
  {2009})}\BibitemShut {NoStop}%
\bibitem [{\citenamefont {Lin}\ \emph {et~al.}(2010)\citenamefont {Lin},
  \citenamefont {Yang}, \citenamefont {Benia}, \citenamefont {Myrach},
  \citenamefont {Yulikov}, \citenamefont {Aumer}, \citenamefont {Brown},
  \citenamefont {Sterrer}, \citenamefont {Bondarchuk}, \citenamefont
  {Kieseritzky}, \citenamefont {Rocker}, \citenamefont {Risse}, \citenamefont
  {Gao}, \citenamefont {Nilius},\ and\ \citenamefont {H.-J.~Freund}}]{Lin}%
  \BibitemOpen
  \bibfield  {author} {\bibinfo {author} {\bibfnamefont {X.}~\bibnamefont
  {Lin}}, \bibinfo {author} {\bibfnamefont {B.}~\bibnamefont {Yang}}, \bibinfo
  {author} {\bibfnamefont {H.~M.}\ \bibnamefont {Benia}}, \bibinfo {author}
  {\bibfnamefont {P.}~\bibnamefont {Myrach}}, \bibinfo {author} {\bibfnamefont
  {M.}~\bibnamefont {Yulikov}}, \bibinfo {author} {\bibfnamefont
  {A.}~\bibnamefont {Aumer}}, \bibinfo {author} {\bibfnamefont {M.~A.}\
  \bibnamefont {Brown}}, \bibinfo {author} {\bibfnamefont {M.}~\bibnamefont
  {Sterrer}}, \bibinfo {author} {\bibfnamefont {O.}~\bibnamefont {Bondarchuk}},
  \bibinfo {author} {\bibfnamefont {E.}~\bibnamefont {Kieseritzky}}, \bibinfo
  {author} {\bibfnamefont {J.}~\bibnamefont {Rocker}}, \bibinfo {author}
  {\bibfnamefont {T.}~\bibnamefont {Risse}}, \bibinfo {author} {\bibfnamefont
  {H.-J.}\ \bibnamefont {Gao}}, \bibinfo {author} {\bibfnamefont
  {N.}~\bibnamefont {Nilius}}, \ and\ \bibinfo {author} {\bibfnamefont
  {J.~A.~C.}\ \bibnamefont {H.-J.~Freund}},\ }\href@noop {} {\ \textbf
  {\bibinfo {volume} {7745}} (\bibinfo {year} {2010})}\BibitemShut {NoStop}%
\bibitem [{\citenamefont {Valeri}\ \emph {et~al.}(2002)\citenamefont {Valeri},
  \citenamefont {Altieri}, \citenamefont {Bona}, \citenamefont {Luches},\ and\
  \citenamefont {Giovanardi}}]{Valeri}%
  \BibitemOpen
  \bibfield  {author} {\bibinfo {author} {\bibfnamefont {S.}~\bibnamefont
  {Valeri}}, \bibinfo {author} {\bibfnamefont {S.}~\bibnamefont {Altieri}},
  \bibinfo {author} {\bibfnamefont {A.~d.}\ \bibnamefont {Bona}}, \bibinfo
  {author} {\bibfnamefont {P.}~\bibnamefont {Luches}}, \ and\ \bibinfo {author}
  {\bibfnamefont {C.~e.}\ \bibnamefont {Giovanardi}},\ }\href@noop {}
  {\bibfield  {journal} {\bibinfo  {journal} {Surf.sci}\ }\textbf {\bibinfo
  {volume} {507}} (\bibinfo {year} {2002})}\BibitemShut {NoStop}%
\bibitem [{\citenamefont {Ferrari}\ \emph {et~al.}(2005)\citenamefont
  {Ferrari}, \citenamefont {Casassa},\ and\ \citenamefont {Pisani}}]{Ferrari}%
  \BibitemOpen
  \bibfield  {author} {\bibinfo {author} {\bibfnamefont {A.~M.}\ \bibnamefont
  {Ferrari}}, \bibinfo {author} {\bibfnamefont {S.}~\bibnamefont {Casassa}}, \
  and\ \bibinfo {author} {\bibfnamefont {C.}~\bibnamefont {Pisani}},\
  }\href@noop {} {\bibfield  {journal} {\bibinfo  {journal} {Phys.Rev.B}\
  }\textbf {\bibinfo {volume} {71}},\ \bibinfo {pages} {155404} (\bibinfo
  {year} {2005})}\BibitemShut {NoStop}%
\bibitem [{\citenamefont {Bieletzki}\ \emph {et~al.}(2010)\citenamefont
  {Bieletzki}, \citenamefont {Hynninen}, \citenamefont {Soini}, \citenamefont
  {Pivetta},\ and\ \citenamefont {Henry}}]{Bieletzki}%
  \BibitemOpen
  \bibfield  {author} {\bibinfo {author} {\bibfnamefont {M.}~\bibnamefont
  {Bieletzki}}, \bibinfo {author} {\bibfnamefont {T.}~\bibnamefont {Hynninen}},
  \bibinfo {author} {\bibfnamefont {T.~M.}\ \bibnamefont {Soini}}, \bibinfo
  {author} {\bibfnamefont {M.}~\bibnamefont {Pivetta}}, \ and\ \bibinfo
  {author} {\bibfnamefont {M.~C. R.~e.}\ \bibnamefont {Henry}},\ }\href@noop {}
  {\bibfield  {journal} {\bibinfo  {journal} {Phys. Chem. Chem Phys}\ }\textbf
  {\bibinfo {volume} {12}},\ \bibinfo {pages} {3203} (\bibinfo {year}
  {2010})}\BibitemShut {NoStop}%
\bibitem [{\citenamefont {Noguera}\ \emph {et~al.}(2010)\citenamefont
  {Noguera}, \citenamefont {Godet},\ and\ \citenamefont
  {Goniakowski}}]{Noguera}%
  \BibitemOpen
  \bibfield  {author} {\bibinfo {author} {\bibfnamefont {C.}~\bibnamefont
  {Noguera}}, \bibinfo {author} {\bibfnamefont {J.}~\bibnamefont {Godet}}, \
  and\ \bibinfo {author} {\bibfnamefont {J.}~\bibnamefont {Goniakowski}},\
  }\href@noop {} {\bibfield  {journal} {\bibinfo  {journal} {Phys. Rev. B}\
  }\textbf {\bibinfo {volume} {81}},\ \bibinfo {pages} {155409} (\bibinfo
  {year} {2010})}\BibitemShut {NoStop}%
\bibitem [{\citenamefont {Schintke}\ \emph {et~al.}(2001)\citenamefont
  {Schintke}, \citenamefont {Messerli}, \citenamefont {Pivetta}, \citenamefont
  {Patthey},\ and\ \citenamefont {Libioulle}}]{Schintke}%
  \BibitemOpen
  \bibfield  {author} {\bibinfo {author} {\bibfnamefont {S.}~\bibnamefont
  {Schintke}}, \bibinfo {author} {\bibfnamefont {S.}~\bibnamefont {Messerli}},
  \bibinfo {author} {\bibfnamefont {M.}~\bibnamefont {Pivetta}}, \bibinfo
  {author} {\bibfnamefont {F.}~\bibnamefont {Patthey}}, \ and\ \bibinfo
  {author} {\bibfnamefont {L.~e.}\ \bibnamefont {Libioulle}},\ }\href@noop {}
  {\ \textbf {\bibinfo {volume} {87}} (\bibinfo {year} {2001})}\BibitemShut
  {NoStop}%
\bibitem [{\citenamefont {Sterrer}\ \emph {et~al.}(2005)\citenamefont
  {Sterrer}, \citenamefont {Fischbach}, \citenamefont {Risse},\ and\
  \citenamefont {Freund}}]{Sterrer}%
  \BibitemOpen
  \bibfield  {author} {\bibinfo {author} {\bibfnamefont {M.}~\bibnamefont
  {Sterrer}}, \bibinfo {author} {\bibfnamefont {E.}~\bibnamefont {Fischbach}},
  \bibinfo {author} {\bibfnamefont {T.}~\bibnamefont {Risse}}, \ and\ \bibinfo
  {author} {\bibfnamefont {H.~J.}\ \bibnamefont {Freund}},\ }\href@noop {}
  {\bibfield  {journal} {\bibinfo  {journal} {Phys. Rev. Lett}\ }\textbf
  {\bibinfo {volume} {94}},\ \bibinfo {pages} {186101} (\bibinfo {year}
  {2005})}\BibitemShut {NoStop}%
\bibitem [{\citenamefont {Cabailh}\ \emph {et~al.}(2011)\citenamefont
  {Cabailh}, \citenamefont {Lazzari}, \citenamefont {Jupille},\ and\
  \citenamefont {L.}}]{Cabailh}%
  \BibitemOpen
  \bibfield  {author} {\bibinfo {author} {\bibfnamefont {G.}~\bibnamefont
  {Cabailh}}, \bibinfo {author} {\bibfnamefont {R.}~\bibnamefont {Lazzari}},
  \bibinfo {author} {\bibfnamefont {S.}~\bibnamefont {Jupille}}, \ and\
  \bibinfo {author} {\bibfnamefont {S.~M.~e.}\ \bibnamefont {L.}},\ }\href@noop
  {} {\bibfield  {journal} {\bibinfo  {journal} {J. Phys. Chem. A.}\ }\textbf
  {\bibinfo {volume} {115}},\ \bibinfo {pages} {7161} (\bibinfo {year}
  {2011})}\BibitemShut {NoStop}%
\bibitem [{\citenamefont {Ouvrard}\ \emph {et~al.}(2011)\citenamefont
  {Ouvrard}, \citenamefont {Niebauer}, \citenamefont {Ghalgaoui}, \citenamefont
  {Barth},\ and\ \citenamefont {Henry}}]{Ouvrard}%
  \BibitemOpen
  \bibfield  {author} {\bibinfo {author} {\bibfnamefont {A.}~\bibnamefont
  {Ouvrard}}, \bibinfo {author} {\bibfnamefont {J.}~\bibnamefont {Niebauer}},
  \bibinfo {author} {\bibfnamefont {A.}~\bibnamefont {Ghalgaoui}}, \bibinfo
  {author} {\bibfnamefont {C.}~\bibnamefont {Barth}}, \ and\ \bibinfo {author}
  {\bibfnamefont {C.~R. e.~a.}\ \bibnamefont {Henry}},\ }\href@noop {}
  {\bibfield  {journal} {\bibinfo  {journal} {J. Phys. Chem. C}\ }\textbf
  {\bibinfo {volume} {115}},\ \bibinfo {pages} {8034} (\bibinfo {year}
  {2011})}\BibitemShut {NoStop}%
\bibitem [{\citenamefont {Baumann}\ \emph {et~al.}(2014)\citenamefont
  {Baumann}, \citenamefont {Rau}, \citenamefont {Loth}, \citenamefont {Lutz},\
  and\ \citenamefont {Heinrich}}]{acsnano}%
  \BibitemOpen
  \bibfield  {author} {\bibinfo {author} {\bibfnamefont {S.}~\bibnamefont
  {Baumann}}, \bibinfo {author} {\bibfnamefont {I.~G.}\ \bibnamefont {Rau}},
  \bibinfo {author} {\bibfnamefont {S.}~\bibnamefont {Loth}}, \bibinfo {author}
  {\bibfnamefont {C.~P.}\ \bibnamefont {Lutz}}, \ and\ \bibinfo {author}
  {\bibfnamefont {A.~J.}\ \bibnamefont {Heinrich}},\ }\href@noop {} {\bibfield
  {journal} {\bibinfo  {journal} {Acs Nano}\ }\textbf {\bibinfo {volume} {8}},\
  \bibinfo {pages} {1739} (\bibinfo {year} {2014})}\BibitemShut {NoStop}%
\bibitem [{\citenamefont {Malashevich}\ \emph {et~al.}(2014)\citenamefont
  {Malashevich}, \citenamefont {Altman},\ and\ \citenamefont
  {Ismail-Beigi}}]{beigi2014}%
  \BibitemOpen
  \bibfield  {author} {\bibinfo {author} {\bibfnamefont {A.}~\bibnamefont
  {Malashevich}}, \bibinfo {author} {\bibfnamefont {E.~I.}\ \bibnamefont
  {Altman}}, \ and\ \bibinfo {author} {\bibfnamefont {S.}~\bibnamefont
  {Ismail-Beigi}},\ }\href@noop {} {\bibfield  {journal} {\bibinfo  {journal}
  {Phy. Rev. B}\ }\textbf {\bibinfo {volume} {90}},\ \bibinfo {pages} {165426}
  (\bibinfo {year} {2014})}\BibitemShut {NoStop}%
\bibitem [{\citenamefont {Stavale}\ \emph {et~al.}(2012)\citenamefont
  {Stavale}, \citenamefont {Shao}, \citenamefont {Nilius}, \citenamefont
  {Freund},\ and\ \citenamefont {Prada}}]{Stavale}%
  \BibitemOpen
  \bibfield  {author} {\bibinfo {author} {\bibfnamefont {F.}~\bibnamefont
  {Stavale}}, \bibinfo {author} {\bibfnamefont {X.}~\bibnamefont {Shao}},
  \bibinfo {author} {\bibfnamefont {N.}~\bibnamefont {Nilius}}, \bibinfo
  {author} {\bibfnamefont {H.-J.}\ \bibnamefont {Freund}}, \ and\ \bibinfo
  {author} {\bibfnamefont {S.}~\bibnamefont {Prada}},\ }\href@noop {}
  {\bibfield  {journal} {\bibinfo  {journal} {J. Am. Chem. Soc}\ }\textbf
  {\bibinfo {volume} {134}},\ \bibinfo {pages} {11380} (\bibinfo {year}
  {2012})}\BibitemShut {NoStop}%
\bibitem [{\citenamefont {Gangopadhyay}()}]{hossein}%
  \BibitemOpen
  \bibfield  {author} {\bibinfo {author} {\bibfnamefont {S.~e.~a.}\
  \bibnamefont {Gangopadhyay}},\ }\href@noop {} {}\ (\bibinfo  {publisher}
  {Under review})\BibitemShut {NoStop}%
\bibitem [{\citenamefont {Markovits}\ \emph {et~al.}(2003)\citenamefont
  {Markovits}, \citenamefont {Paniagua}, \citenamefont {Lopez}, \citenamefont
  {Minot},\ and\ \citenamefont {Illas}}]{illasprb}%
  \BibitemOpen
  \bibfield  {author} {\bibinfo {author} {\bibfnamefont {A.}~\bibnamefont
  {Markovits}}, \bibinfo {author} {\bibfnamefont {J.~C.}\ \bibnamefont
  {Paniagua}}, \bibinfo {author} {\bibfnamefont {N.}~\bibnamefont {Lopez}},
  \bibinfo {author} {\bibfnamefont {C.}~\bibnamefont {Minot}}, \ and\ \bibinfo
  {author} {\bibfnamefont {F.}~\bibnamefont {Illas}},\ }\href@noop {}
  {\bibfield  {journal} {\bibinfo  {journal} {Phys. Rev. B}\ }\textbf {\bibinfo
  {volume} {67}},\ \bibinfo {pages} {115417} (\bibinfo {year}
  {2003})}\BibitemShut {NoStop}%
\bibitem [{\citenamefont {Rau}(2014)}]{science2014}%
  \BibitemOpen
  \bibfield  {author} {\bibinfo {author} {\bibfnamefont {I.~G.}\ \bibnamefont
  {Rau}},\ }\href@noop {} {\bibfield  {journal} {\bibinfo  {journal} {Reaching
  the magnetic anisotropy limit of a}\ }\textbf {\bibinfo {volume} {344}},\
  \bibinfo {pages} {988} (\bibinfo {year} {2014})}\BibitemShut {NoStop}%
\bibitem [{\citenamefont {Albertini}\ \emph {et~al.}(2015)\citenamefont
  {Albertini}, \citenamefont {Liu},\ and\ \citenamefont {Jones}}]{oliver}%
  \BibitemOpen
  \bibfield  {author} {\bibinfo {author} {\bibfnamefont {O.}~\bibnamefont
  {Albertini}}, \bibinfo {author} {\bibfnamefont {A.~Y.}\ \bibnamefont {Liu}},
  \ and\ \bibinfo {author} {\bibfnamefont {B.~A.}\ \bibnamefont {Jones}},\
  }\href@noop {} {\bibfield  {journal} {\bibinfo  {journal} {Phys. Rev. B}\
  }\textbf {\bibinfo {volume} {91}},\ \bibinfo {pages} {214423} (\bibinfo
  {year} {2015})}\BibitemShut {NoStop}%
\bibitem [{\citenamefont {Lin}\ and\ \citenamefont {Jones}(2011)}]{barbara}%
  \BibitemOpen
  \bibfield  {author} {\bibinfo {author} {\bibfnamefont {C.-Y.}\ \bibnamefont
  {Lin}}\ and\ \bibinfo {author} {\bibfnamefont {B.~A.}\ \bibnamefont
  {Jones}},\ }\href@noop {} {\bibfield  {journal} {\bibinfo  {journal} {Phys.
  Rev. B}\ }\textbf {\bibinfo {volume} {83}},\ \bibinfo {pages} {014413}
  (\bibinfo {year} {2011})}\BibitemShut {NoStop}%
\bibitem [{\citenamefont {Kokalj}(2003)}]{xcrysden}%
  \BibitemOpen
  \bibfield  {author} {\bibinfo {author} {\bibfnamefont {A.}~\bibnamefont
  {Kokalj}},\ }\href@noop {} {\bibfield  {journal} {\bibinfo  {journal} {Comp.
  Mater. Sci}\ }\textbf {\bibinfo {volume} {28}},\ \bibinfo {pages} {155}
  (\bibinfo {year} {2003})}\BibitemShut {NoStop}%
\bibitem [{\citenamefont {Bryan}\ \emph {et~al.}(2012)\citenamefont {Bryan},
  \citenamefont {Merrill}, \citenamefont {Reiff}, \citenamefont {Fettinger},\
  and\ \citenamefont {Power}}]{stm19}%
  \BibitemOpen
  \bibfield  {author} {\bibinfo {author} {\bibfnamefont {A.~M.}\ \bibnamefont
  {Bryan}}, \bibinfo {author} {\bibfnamefont {W.~A.}\ \bibnamefont {Merrill}},
  \bibinfo {author} {\bibfnamefont {W.~M.}\ \bibnamefont {Reiff}}, \bibinfo
  {author} {\bibfnamefont {J.~C.}\ \bibnamefont {Fettinger}}, \ and\ \bibinfo
  {author} {\bibfnamefont {P.~P.~S.}\ \bibnamefont {Power}},\ }\href@noop {}
  {\bibfield  {journal} {\bibinfo  {journal} {Inorg. Chem}\ }\textbf {\bibinfo
  {volume} {51}},\ \bibinfo {pages} {3366} (\bibinfo {year}
  {2012})}\BibitemShut {NoStop}%
\bibitem [{\citenamefont {Loth}\ \emph {et~al.}(2010)\citenamefont {Loth},
  \citenamefont {Etzkorn}, \citenamefont {Lutz}, \citenamefont {Eigler},\ and\
  \citenamefont {Heinrich}}]{stm29}%
  \BibitemOpen
  \bibfield  {author} {\bibinfo {author} {\bibfnamefont {S.}~\bibnamefont
  {Loth}}, \bibinfo {author} {\bibfnamefont {M.}~\bibnamefont {Etzkorn}},
  \bibinfo {author} {\bibfnamefont {C.~P.}\ \bibnamefont {Lutz}}, \bibinfo
  {author} {\bibfnamefont {D.~M.}\ \bibnamefont {Eigler}}, \ and\ \bibinfo
  {author} {\bibfnamefont {A.~J.}\ \bibnamefont {Heinrich}},\ }\href@noop {}
  {\bibfield  {journal} {\bibinfo  {journal} {Science}\ }\textbf {\bibinfo
  {volume} {329}},\ \bibinfo {pages} {1628} (\bibinfo {year}
  {2010})}\BibitemShut {NoStop}%
\bibitem [{\citenamefont {Groot}(2001)}]{stm30}%
  \BibitemOpen
  \bibfield  {author} {\bibinfo {author} {\bibfnamefont {F.~d.}\ \bibnamefont
  {Groot}},\ }\href@noop {} {\bibfield  {journal} {\bibinfo  {journal} {Chem.
  Rev}\ }\textbf {\bibinfo {volume} {101}},\ \bibinfo {pages} {1779} (\bibinfo
  {year} {2001})}\BibitemShut {NoStop}%
\bibitem [{\citenamefont {Degroot}\ \emph {et~al.}(1993)\citenamefont
  {Degroot}, \citenamefont {Abbate}, \citenamefont {Vanelp}, \citenamefont
  {Sawatzky},\ and\ \citenamefont {Ma}}]{stm31}%
  \BibitemOpen
  \bibfield  {author} {\bibinfo {author} {\bibfnamefont {F.~M. F.~d.}\
  \bibnamefont {Degroot}}, \bibinfo {author} {\bibfnamefont {M.}~\bibnamefont
  {Abbate}}, \bibinfo {author} {\bibfnamefont {J.}~\bibnamefont {Vanelp}},
  \bibinfo {author} {\bibfnamefont {G.~A.}\ \bibnamefont {Sawatzky}}, \ and\
  \bibinfo {author} {\bibfnamefont {Y.~J.}\ \bibnamefont {Ma}},\ }\href@noop {}
  {\bibfield  {journal} {\bibinfo  {journal} {J. Phys. Condens. Matter}\
  }\textbf {\bibinfo {volume} {5}},\ \bibinfo {pages} {2277} (\bibinfo {year}
  {1993})}\BibitemShut {NoStop}%
\bibitem [{\citenamefont {Gambardella}\ \emph {et~al.}(2002)\citenamefont
  {Gambardella}, \citenamefont {Dhesi}, \citenamefont {Gardonio}, \citenamefont
  {Grazioli},\ and\ \citenamefont {Ohresser}}]{stm32}%
  \BibitemOpen
  \bibfield  {author} {\bibinfo {author} {\bibfnamefont {P.}~\bibnamefont
  {Gambardella}}, \bibinfo {author} {\bibfnamefont {S.~S.}\ \bibnamefont
  {Dhesi}}, \bibinfo {author} {\bibfnamefont {S.}~\bibnamefont {Gardonio}},
  \bibinfo {author} {\bibfnamefont {C.}~\bibnamefont {Grazioli}}, \ and\
  \bibinfo {author} {\bibfnamefont {P.}~\bibnamefont {Ohresser}},\ }\href@noop
  {} {\bibfield  {journal} {\bibinfo  {journal} {Phys. Rev. Lett}\ }\textbf
  {\bibinfo {volume} {88}},\ \bibinfo {pages} {047202} (\bibinfo {year}
  {2002})}\BibitemShut {NoStop}%
\bibitem [{\citenamefont {Cococcioni}\ and\ \citenamefont
  {Gironcoli}(2005)}]{U_qe}%
  \BibitemOpen
  \bibfield  {author} {\bibinfo {author} {\bibfnamefont {M.}~\bibnamefont
  {Cococcioni}}\ and\ \bibinfo {author} {\bibfnamefont {S.~d.}\ \bibnamefont
  {Gironcoli}},\ }\href@noop {} {\bibfield  {journal} {\bibinfo  {journal}
  {Phys. Rev. B}\ }\textbf {\bibinfo {volume} {71}},\ \bibinfo {pages} {035105}
  (\bibinfo {year} {2005})}\BibitemShut {NoStop}%
\bibitem [{\citenamefont {Zhou}\ and\ \citenamefont {Ju}(2013)}]{Zhou}%
  \BibitemOpen
  \bibfield  {author} {\bibinfo {author} {\bibfnamefont {N.}~\bibnamefont
  {Zhou}}\ and\ \bibinfo {author} {\bibfnamefont {D.~A.}\ \bibnamefont {Ju}},\
  }\href@noop {} {\bibfield  {journal} {\bibinfo  {journal} {J. Electrochem.
  Soc}\ }\textbf {\bibinfo {volume} {160}},\ \bibinfo {pages} {A1863} (\bibinfo
  {year} {2013})}\BibitemShut {NoStop}%
\bibitem [{\citenamefont {Zhang}\ \emph {et~al.}(2013)\citenamefont {Zhang},
  \citenamefont {Yin}, \citenamefont {Xia}, \citenamefont {Zhang},\ and\
  \citenamefont {Liu}}]{jap}%
  \BibitemOpen
  \bibfield  {author} {\bibinfo {author} {\bibfnamefont {T.}~\bibnamefont
  {Zhang}}, \bibinfo {author} {\bibfnamefont {J.}~\bibnamefont {Yin}}, \bibinfo
  {author} {\bibfnamefont {Y.}~\bibnamefont {Xia}}, \bibinfo {author}
  {\bibfnamefont {W.}~\bibnamefont {Zhang}}, \ and\ \bibinfo {author}
  {\bibfnamefont {Z.}~\bibnamefont {Liu}},\ }\href@noop {} {\bibfield
  {journal} {\bibinfo  {journal} {J. Appl. Phys}\ }\textbf {\bibinfo {volume}
  {114}},\ \bibinfo {pages} {134301} (\bibinfo {year} {2013})}\BibitemShut
  {NoStop}%
\bibitem [{\citenamefont {Shin}\ \emph {et~al.}(2010)\citenamefont {Shin},
  \citenamefont {Jung}, \citenamefont {Motobayashi}, \citenamefont
  {Yanagisawa},\ and\ \citenamefont {Morikawa}}]{nature}%
  \BibitemOpen
  \bibfield  {author} {\bibinfo {author} {\bibfnamefont {H.-J.}\ \bibnamefont
  {Shin}}, \bibinfo {author} {\bibfnamefont {J.}~\bibnamefont {Jung}}, \bibinfo
  {author} {\bibfnamefont {K.}~\bibnamefont {Motobayashi}}, \bibinfo {author}
  {\bibfnamefont {S.}~\bibnamefont {Yanagisawa}}, \ and\ \bibinfo {author}
  {\bibfnamefont {Y.}~\bibnamefont {Morikawa}},\ }\href@noop {} {\bibfield
  {journal} {\bibinfo  {journal} {Nat. Mater}\ }\textbf {\bibinfo {volume}
  {9}},\ \bibinfo {pages} {442} (\bibinfo {year} {2010})}\BibitemShut {NoStop}%
\bibitem [{\citenamefont {Perdew}\ \emph {et~al.}(1996)\citenamefont {Perdew},
  \citenamefont {Burke},\ and\ \citenamefont {Ernzerhof}}]{pbe}%
  \BibitemOpen
  \bibfield  {author} {\bibinfo {author} {\bibfnamefont {J.~P.}\ \bibnamefont
  {Perdew}}, \bibinfo {author} {\bibfnamefont {K.}~\bibnamefont {Burke}}, \
  and\ \bibinfo {author} {\bibfnamefont {M.}~\bibnamefont {Ernzerhof}},\
  }\href@noop {} {\bibfield  {journal} {\bibinfo  {journal} {Phys. Rev. Lett}\
  }\textbf {\bibinfo {volume} {77}},\ \bibinfo {pages} {3865} (\bibinfo {year}
  {1996})}\BibitemShut {NoStop}%
\bibitem [{\citenamefont {Vanderbilt}(1990)}]{ultrasoft}%
  \BibitemOpen
  \bibfield  {author} {\bibinfo {author} {\bibfnamefont {D.}~\bibnamefont
  {Vanderbilt}},\ }\href@noop {} {\bibfield  {journal} {\bibinfo  {journal}
  {Phys. Rev. B}\ }\textbf {\bibinfo {volume} {41}},\ \bibinfo {pages} {7892}
  (\bibinfo {year} {1990})}\BibitemShut {NoStop}%
\bibitem [{\citenamefont {Giannozzi}\ \emph {et~al.}(2009)\citenamefont
  {Giannozzi}, \citenamefont {Baroni}, \citenamefont {Bonini}, \citenamefont
  {Calandra},\ and\ \citenamefont {Car}}]{qe}%
  \BibitemOpen
  \bibfield  {author} {\bibinfo {author} {\bibfnamefont {P.}~\bibnamefont
  {Giannozzi}}, \bibinfo {author} {\bibfnamefont {S.}~\bibnamefont {Baroni}},
  \bibinfo {author} {\bibfnamefont {N.}~\bibnamefont {Bonini}}, \bibinfo
  {author} {\bibfnamefont {M.}~\bibnamefont {Calandra}}, \ and\ \bibinfo
  {author} {\bibfnamefont {R.~Q.~E.}\ \bibnamefont {Car}},\ }\href@noop {}
  {\bibfield  {journal} {\bibinfo  {journal} {J. Phys.: Condens. Matter}\
  }\textbf {\bibinfo {volume} {21}},\ \bibinfo {pages} {395502} (\bibinfo
  {year} {2009})}\BibitemShut {NoStop}%
\bibitem [{\citenamefont {Blaha}\ \emph {et~al.}()\citenamefont {Blaha},
  \citenamefont {Schwarz}, \citenamefont {Madsen}, \citenamefont {Kvasicka},\
  and\ \citenamefont {Luitz}}]{wien2k}%
  \BibitemOpen
  \bibfield  {author} {\bibinfo {author} {\bibfnamefont {P.}~\bibnamefont
  {Blaha}}, \bibinfo {author} {\bibfnamefont {K.}~\bibnamefont {Schwarz}},
  \bibinfo {author} {\bibfnamefont {G.}~\bibnamefont {Madsen}}, \bibinfo
  {author} {\bibfnamefont {D.}~\bibnamefont {Kvasicka}}, \ and\ \bibinfo
  {author} {\bibfnamefont {J.~W.}\ \bibnamefont {Luitz}},\ }\href@noop {} {\
  \textbf {\bibinfo {volume} {2001}}}\BibitemShut {NoStop}%
\bibitem [{\citenamefont {Madsen}\ and\ \citenamefont {Nov\'ak}(2005)}]{epl}%
  \BibitemOpen
  \bibfield  {author} {\bibinfo {author} {\bibfnamefont {G.~K.~H.}\
  \bibnamefont {Madsen}}\ and\ \bibinfo {author} {\bibfnamefont
  {P.}~\bibnamefont {Nov\'ak}},\ }\href@noop {} {\bibfield  {journal} {\bibinfo
   {journal} {Euro. Phys. Lett}\ }\textbf {\bibinfo {volume} {69}},\ \bibinfo
  {pages} {777} (\bibinfo {year} {2005})}\BibitemShut {NoStop}%
\end{thebibliography}%

\end{document}